\documentclass[a4paper,11pt]{article}
\usepackage{jheppub}

\pdfoutput=1
\usepackage{hyperref}
\usepackage{graphicx}
\usepackage[utf8]{inputenc}
\usepackage{xspace}
\usepackage{amsmath,amssymb}
\usepackage{bbm}
\usepackage{xcolor}
\usepackage{booktabs}
\usepackage{subcaption}
\usepackage{mathtools}
\usepackage[ruled,norelsize]{algorithm2e}
\usepackage{empheq}
\usepackage{cancel}
\usepackage{multirow}
\usepackage{slashed}
\usepackage{feynmp}
\usepackage{feynmp-auto}
\usepackage{tikz}
\usetikzlibrary{decorations.pathmorphing}
\usetikzlibrary{decorations.markings}
\usetikzlibrary{positioning, shapes, snakes, arrows}
\tikzset{
	quark/.style={postaction={decorate},
		decoration={markings,mark=at position .5 with {\arrow[#1]{latex}}}},
	scalar/.style={dashed,postaction={decorate},
		decoration={markings,mark=at position .5 with {\arrow[#1]{latex}}}},
	gluon/.style={decorate,
		decoration={coil,amplitude=2pt, segment length=2pt,  pre length=.1cm, post length=.1cm}},
	boson/.style={-latex,decorate, decoration={snake, segment length=4pt, amplitude=1.8pt, pre length=.1cm, post length=.25cm}},
	photon/.style={decorate, decoration={snake, segment length=4pt, amplitude=1.8pt,  pre length=.1cm, post length=.1cm}},
	dphoton/.style={decorate, decoration={snake, segment length=4pt, amplitude=1.8pt,  pre length=.1cm, post length=.25cm},-latex}
}

\newcommand{\ep}{\epsilon}
\newcommand{\zc}{z_{\mathrm{cut}}}

\newcommand{\nn}{\nonumber}
\newcommand{\sd}{d}

\DeclareFontFamily{U}{rcjhbltx}{}
\DeclareFontShape{U}{rcjhbltx}{m}{n}{<->rcjhbltx}{}
\title{Collinear fragmentation at NNLL: generating functionals,
  groomed correlators and angularities}

\author[a]{Melissa van Beekveld,}
\author[b,c]{Mrinal Dasgupta,}
\author[b]{Basem Kamal El-Menoufi,}
\author[a]{Jack Helliwell,}
\author[c]{and Pier Francesco Monni}

\affiliation[a]{Rudolf Peierls Centre for Theoretical Physics, Clarendon
  Laboratory, Parks Road, University of Oxford, Oxford OX1 3PU, UK}
\affiliation[b]{Lancaster-Manchester-Sheffield Consortium for
  Fundamental Physics, Department of Physics \& Astronomy, University of
  Manchester, Manchester M13 9PL, United Kingdom}
\affiliation[c]{CERN, Theoretical Physics Department, CH-1211 Geneva
  23, Switzerland}

\emailAdd{melissa.vanbeekveld@physics.ox.ac.uk}
\emailAdd{mrinal.dasgupta@manchester.ac.uk}
\emailAdd{basem.el-menoufi@manchester.ac.uk}
\emailAdd{jack.helliwell@physics.ox.ac.uk}
\emailAdd{pier.monni@cern.ch}

\preprint{CERN-TH-2023-143, OUTP-23-08P}

\abstract{Jet calculus offers a unique mathematical technique to
  bridge the area of QCD resummation with Monte Carlo parton showers.
  With the ultimate goal of constructing next-to-next-to-leading
  logarithmic (NNLL) parton showers we study, using the language of
  generating functionals, the collinear fragmentation of final-state
  partons. In particular, we focus on the definition and calculation
  of the Sudakov form factor, which physically describes the
  no-emission probability in an ordered branching process.
  We review recent results for quark jets and compute the Sudakov form
  factor for the collinear fragmentation of gluon jets at NNLL.
  The NNLL corrections are encoded in a $z$ dependent two-loop
  anomalous dimension ${\mathcal B}_2(z)$, with $z$ being a suitably
  defined longitudinal momentum fraction. This is obtained from the
  integration of the relevant $1\to 3$ collinear splitting kernels
  combined with the one-loop corrections to the $1\to 2$ counterpart.
  This work provides the missing ingredients to extend the methods of
  jet calculus in the collinear limit to NNLL and gives an important element
  of the next generation of NNLL parton shower algorithms.
  As an application we derive new NNLL results for both the
  fractional moments of energy-energy correlation $FC_x$ and the
  angularities $\lambda_x$ measured on mMDT/Soft-Drop ($\beta=0$)
  groomed jets.}

\begin{document}

\setlength{\parskip}{0pt}
\maketitle
\flushbottom

\section{Introduction and motivation}
\label{sec:intro}
In this paper and in a forthcoming article~\cite{vanBeekveld:2023?}, we
initiate a formulation of the methods of jet
calculus~\cite{Konishi:1979cb,Bassetto:1983mvz,Dokshitzer:1991wu,Ellis:1996mzs}
beyond the NLL order to describe the dynamics of collinear
fragmentation.
Jet calculus techniques, and specifically the generating functional
method, stand out as a mathematical language to formulate self-similar
branching processes. For this reason, they are particularly suitable
to bridge the field of QCD resummation to that of parton shower Monte
Carlo algorithms. This is an important direction to pursue in the context of
improving the logarithmic accuracy of such algorithms for QCD
phenomenology and is receiving widespread attention in the collider
physics context (see e.g. Ref.~\cite{Campbell:2022qmc} for an
overview).
 
 More specifically this work addresses the following points: 
 \begin{enumerate}
 \item A large class of QCD resummations cannot be handled using
   purely analytic techniques since they do not admit a closed
   analytic form. This can be either due to the non-linear nature of
   the evolution equations (like for instance the resummation of microjet
   observables~\cite{Dasgupta:2014yra,Kang:2016mcy},
   fragmentation~\cite{Jain:2011xz,Chang:2013rca,Ritzmann:2014mka,Elder:2017bkd,Chen:2020adz,Neill:2020bwv,Chen:2021gdk,Neill:2021std,Chen:2022jhb,Chen:2022muj,Chen:2023zlx}
   or non-global
   observables~\cite{Dasgupta:2001sh,Banfi:2002hw,Hatta:2013iba,Caron-Huot:2015bja,Larkoski:2015zka,Neill:2016stq,AngelesMartinez:2018cfz,Hatta:2020wre,Banfi:2021xzn,Banfi:2021owj,Becher:2023vrh})
   or to the high complexity of the observable which may not easily
   factorise in a suitable conjugate space (like for instance some
   complex event shapes or jet resolution
   parameters~\cite{Banfi:2001bz,Banfi:2014sua,Banfi:2016zlc}). In
   such cases it is necessary to formulate the resummation in a way
   that can be solved accurately using numerical methods. The methods
   of jet
   calculus~\cite{Konishi:1979cb,Bassetto:1983mvz,Dokshitzer:1991wu,Ellis:1996mzs},
   notably Generating Functionals (GFs), offer a powerful mathematical
   tool to describe the resummation of logarithmic radiative
   corrections in collider observables, and constitute one avenue to
   achieve a numerical solution via Monte Carlo techniques.
 \item The recent development of new techniques to construct parton-shower algorithms with demonstrable higher logarithmic accuracy
   (see
   e.g. Refs.~\cite{Li:2016yez,Hoche:2017hno,Dasgupta:2018nvj,Dulat:2018vuy,Bewick:2019rbu,Forshaw:2019ver,Dasgupta:2020fwr,Forshaw:2020wrq,Hamilton:2020rcu,Nagy:2020rmk,Nagy:2020dvz,Hamilton:2021dyz,Karlberg:2021kwr,Herren:2022jej,vanBeekveld:2022zhl,vanBeekveld:2022ukn,Hamilton:2023dwb,vanBeekveld:2023lfu,Assi:2023rbu,Ravasio:2023anp})
   offers an opportunity to further increase systematically their
   precision.
   The connection between these new developments in the area of parton
   showers and resummations is based on a well-defined set of accuracy
   criteria that are based directly on QCD dynamics
   (i.e. approximating correctly the real and virtual matrix elements
   in specific kinematical limits). Recent parton-shower
   algorithms~\cite{Dasgupta:2020fwr,Hamilton:2020rcu,Hamilton:2021dyz,Karlberg:2021kwr,vanBeekveld:2022zhl,vanBeekveld:2022ukn,Hamilton:2023dwb,vanBeekveld:2023lfu,Ravasio:2023anp}
   are constructed in such a way that the above criteria are
   satisfied, hence achieving NLL accuracy for broad classes of
   observable at once.
  GFs help us push this important correspondence to a deeper level, by
  deriving evolution equations which provide an analytic connection
  between parton showers and all-order resummations. As a result, one
  can unify parton showers and resummations within a single framework,
  which allows for systematic developments in both areas
  simultaneously.
\end{enumerate}

In this work we focus on the class of observables which feature only
collinear sensitivity, and on the description of final state
fragmentation, leaving the extension to initial state radiation to
future work.%
\footnote{A related description of the evolution of soft radiation
  away from the collinear limit using the same mathematical language
  was discussed in Refs.~\cite{Banfi:2021xzn,Banfi:2021owj} in the
  context of non-global resummations in the large$-N_c$ limit.}
  Such observables are by definition only sensitive to non-soft
  small-angle emissions, where the longitudinal momentum fraction $z$
  of the splitting satisfies $z\sim (1-z) \sim {\cal O}(1)$. In this
  regime, the relative angle and normalised transverse momentum of
  each splitting are commensurate, and one can expand one about the
  other systematically. Examples of observables of this kind are moments of energy-energy correlations~\cite{Banfi:2004yd} or
  angularities~\cite{Berger:2003iw} measured on mMDT/soft-drop
  ($\beta=0$) groomed jets~\cite{Dasgupta:2013ihk,Larkoski:2014wba},
  which we will study here, or fragmentation
  functions~\cite{Gribov:1972ri,Dokshitzer:1977sg,Altarelli:1977zs,Furmanski:1980cm,Curci:1980uw}.

Here we calculate the analytical ingredients needed in the formulation
of NNLL resummation within the GFs method. As we will demonstrate in
forthcoming work~\cite{vanBeekveld:2023?}, these ingredients can then
also be used in Monte Carlo algorithms to correctly include virtual
corrections,\footnote{Specifically, by unitary approximation we mean
  that in a parton shower virtual corrections simply amount to (minus)
  the integral of the real (soft and/or collinear) matrix
  element. This implies a unitary effect on the total cross section.}
which are not captured by the shower's unitary approximation.
For this reason, ingredients of this type (such as the
soft physical coupling
scheme~\cite{Catani:1990rr,Banfi:2018mcq,Catani:2019rvy}) have to be
computed in dimensional regularisation.

The aforementioned analytical ingredients amount to the anomalous
dimensions defining the NNLL Sudakov form factor, which encodes the
no-emission probability in the branching process, and is the back bone
of both the GFs method and parton-shower algorithms.
After reviewing the recent work of Ref.~\cite{Dasgupta:2021hbh}, which
provides the NNLL Sudakov for quark jets, here we frame these results
in the context of the GFs method and furthermore define and compute
the NNLL Sudakov form factor for the case of gluon fragmentation. We
also discuss applications of our results in the context of QCD
resummation, by deriving new NNLL accurate results for specific
groomed jet observables related closely to those measured at the LHC.
For these observables, the evolution equations can be solved in closed
form and allows us to derive analytic results.

The layout of the paper is as follows. In Sec.~\ref{sec:GFs} we
introduce the Generating Functionals method as a way to understand the
role of such analytic ingredients in a Monte Carlo calculation. We
then discuss the extension to NNLL order in the collinear limit and
define the NNLL Sudakov form factor and the anomalous dimension
$\mathcal{B}_2(z)$.
In Secs.~\ref{sect:defs}-\ref{sect:b2extraction} we calculate
$\mathcal{B}^g_2(z)$ for gluon jets, which complements previous
studies done for quark jets and discuss the important differences
between these two cases.
In Sec.~\ref{sec:applications} we provide concrete applications to
collider phenomenology by deriving new results for the NNLL
resummation of moments of energy-energy
correlation~\cite{Banfi:2004yd} and angularities~\cite{Berger:2003iw}
measured on groomed jets with the mMDT/Soft-Drop
procedure~\cite{Dasgupta:2013ihk,Larkoski:2014wba}.
We conclude in Sec.~\ref{sec:conclusions} and provide a
number of appendices with additional technical details about the
method and the calculations performed in this paper.
The results for the $\mathcal{B}_2(z)$ anomalous dimensions for quarks
and gluons are given in electronic format with the arXiv submission of
this article.

\section{Generating functionals for collinear fragmentation}

\label{sec:GFs}
In this section we introduce the GF method to resum logarithmic
corrections of collinear nature. We start by reviewing the NLL case
and then we discuss the extension to NNLL.

\subsection{Review of NLL resummation}
We consider the collinear fragmentation of a parton of initial energy
$E$ resolved at an angular resolution $\theta\ll 1$. The all-order
resummation at NLL resums single logarithms $L = -\ln \theta$ of the
form $\alpha_s^nL^n$. At this order, the fragmentation can be
formulated as a branching process in which emissions are strongly
ordered in angle.\footnote{Angular ordering ensures the full coverage
  of the relevant phase space at NLL. Beyond NLL, emissions can have
  commensurate angles and therefore the precise definition of angular
  ordering (i.e. with respect to a specific reference direction) has
  to be specified in order to guarantee the coverage of the full phase
  space.}
  At NNLL we aim at resumming corrections of order
  $\alpha_s^n L ^{n-1}$, where now emissions can be unordered and have
  commensurate angles.
To start, we therefore define an angular resolution scale $t$
(evolution time) as
\begin{equation}\label{eq:time}
t_i = 
\int_{\theta_i^2}^1\frac{d\theta^2}{\theta^2}\,\frac{\alpha_s(E_p^2g^2(z)\,\theta^2)}{2\pi}\,,
\end{equation}
where $E_p$ is the energy of the parton branching at angular
resolution $t_i$.
In Eq.~\eqref{eq:time} $\alpha_s$ is the $\overline{\rm MS}$ coupling,
and $g(z)$ is a function of the longitudinal momentum fraction $1-z$
carried by the $i$-th emission.
At NLL the precise form of $g(z)$ is irrelevant and one can set
$g(z)=1$ as well as $E_p=E$ (see for example
Ref.~\cite{Dasgupta:2014yra}).
In what follows, the use of the $\beta(\alpha_s)$ function that drives
the running of $\alpha_s$ at either one-loop or two-loop order,
depending on whether the target accuracy is NLL or NNLL is understood.

We introduce the GF $G_f(x,t)$, which encodes the probability of
resolving a fixed number of emissions in the time-like fragmentation
of an initial parton of flavour $f\in \{q,g\}$ and momentum fraction
$x$ (i.e. $E_p\equiv x\, E$) below a resolution angle
set by an initial evolution time $t$.
The GFs are defined in such a way that the probability of exclusively
resolving $m$ partons in the collinear fragmentation of a parton of
flavour $f\in \{q,g\}$ and momentum fraction $x$ starting at an
evolution time $t$ is given by~\cite{Konishi:1979cb,Dokshitzer:1991wu}
\begin{equation}\label{eq:prob-quark}
  \int dP^{(f)}_m = \left. \frac{1}{m!}\frac{\delta^m}{\delta
      u^m}\, G_f(x,t) \right|_{\{u\}=0}\hspace{-0.8cm}.
\end{equation}
The quantity $u\equiv u(x,t;f)$ is the \textit{probing function} and has the role of
tagging a real emission in the functional derivative of $G_f$.
Eq.~\eqref{eq:prob-quark} can be taken as the definition of the
generating functional $G_f$.
The evolution of the GFs with the resolution scale $t$ is described,
at NLL, by the coupled system of integral
equations~\cite{Dasgupta:2014yra}\footnote{An equivalent differential
  form can be easily obtained by dividing $G_f$ by $\Delta_f$ and
  subsequently taking the $t$ derivative. Note that
  Ref.~\cite{Dasgupta:2014yra} adopts a slightly different definition
  of the GFs which in this reference describe the fragmentation of a
  parton from a starting time $t=0$ down to a resolution angle set by
  the time $t$. This convention is complementary to the one adopted in
  this paper, but their difference is irrelevant at the level of
  physical results.}
\begin{align}\label{eq:GF-NLL}
  G_q(x,t) &= u\, \Delta_q(t) + \int_{t}^{t_{0}}\ d t' \int_{z_0}^{1-z_0} d z\,
             P_{qq}(z)\,G_q(x\,z,t') \,G_g(x\,(1-z),t') \frac{\Delta_q(t)}{\Delta_q(t')}\,,\notag\\
G_g(x,t) &= u\, \Delta_g(t) + \int_{t}^{t_0} d t' \int_{z_0}^{1-z_0} d z
  \,\bigg[ P_{gg}(z)\,G_g(x\,z,t')
  \,G_g(x\,(1-z),t') \notag\\
&+  P_{qg}(z)\,G_q(x\,z,t')
  \,G_q(x\,(1-z),t') \bigg]\frac{\Delta_g(t)}{\Delta_g(t')} \,,
\end{align}
where $t_0$ is a collinear cutoff at which the evolution stops, while
$z_0$ is an infrared cutoff on the energy fraction of each emission,
which must be taken to zero in the calculation of an IRC safe
observable (unlike for $z_0$, the value of $t_0$ is bounded by the
presence of the Landau pole). The standard leading-order splitting
functions are given in Appendix~\ref{app:SF}.
The above set of equations can be used to derive collinear
resummations with NLL (single logarithmic) accuracy for final state
radiation.
The Sudakov form factors $\Delta_f$ describe the no-emission
probability and can be derived by imposing the unitarity condition
\begin{equation}\label{eq:unitarity}
\left.G_f(x,t)\right|_{u=1} = 1\,,
\end{equation}
from which one obtains
\begin{align}\label{eq:sudakov-nll}
\ln \Delta_q(t) &= - \int_t^{t_0} d t' \int_{z_0}^{1-z_0}d z\,
  P_{qq}(z)\,,\\
\ln \Delta_g(t) &= - \int_t^{t_0} d t' \int_{z_0}^{1-z_0}d z
  \left(P_{gg}(z)+P_{qg}(z)\right)\,.
\end{align}
An obvious boundary condition is then $G_f(x,t_0)=u$, indicating a
$100\%$ probability of finding a parton $f$ with momentum $x$.
With the above definition of the GFs, the resummed distribution (or
equivalently cumulative distribution) $d\sigma^{(f)}$ for a given
observable $v=V(\{k_i\})$ (with $\{k_i\}$ denoting the final state
momenta produced in the fragmentation of a jet of initial flavour $f$)
has the general form
\begin{align}\label{eq:dsigma}
  d\sigma^{(f)} = \sigma_0 \,C(\alpha_s) \otimes  J^{(f)}(\alpha_s, v)\,,
\end{align}
where $\sigma_0$ is the Born cross section for the hard process under
consideration.
The jet distribution $ J^{(f)}$ is simply obtained by integrating
Eq.~\eqref{eq:prob-quark} with the measurement function of the
observable for any final-state multiplicity $m$, that is
\begin{equation}
 J^{(f)}(\alpha_s, v) = \sum_{m=1}^{\infty} \int \,d P^{(f)}_m\, \delta(v-V(\{k\}_m))\,.
\end{equation}
The $\otimes$ operation is observable dependent. It is usually a
regular product, but for some specific observables (e.g.~fragmentation
functions) it can take the form of a convolution over the longitudinal
momentum fraction.
The process- and observable-dependent matching coefficient
$C(\alpha_s)$ admits a fixed-order perturbative expansion in powers of
the strong coupling constant $C(\alpha_s)=1+{\cal O}(\alpha_s)$, and
it accounts for constant terms stemming from the matching of the jet
distribution (defined by the GFs evolution equation) to the fixed
order QCD calculation in the limit $v\to 0$.
Specifically, at NNLL $C(\alpha_s)$ is required at the one-loop order,
which entails the difference between the full ${\cal O}(\alpha_s)$ QCD
calculation in the logarithmic limit (i.e. $v\to 0$) for the
observable $v$ and the expansion of the jet function at the same
order.

\subsection{Extension to the NNLL case}
The extension of the above formulation to NNLL order requires the
generalisation of the r.h.s. of the evolution
equations~\eqref{eq:GF-NLL} to ${\cal O}(\alpha_s^2)$.
To understand what the calculation entails, 
we follow an analogous derivation of that presented in Refs.~\cite{Banfi:2021xzn,Banfi:2021owj} in the context of soft-gluon evolution in the planar limit. 
We observe that
Eqs.~\eqref{eq:GF-NLL} describe the fragmentation by a sequence of
angular ordered branchings. The resulting emission probability is the
iteration of $1\to 2$ splitting kernels, an independent emission
pattern, which correctly describes the NLL strongly ordered regime,
i.e. $t_i \ll t_{i+1}$. At NNLL one needs to account for unordered
corrections, i.e. $t_{i}\sim t_{i+1}$, which are described by the full
set of $1\to 3$ splitting
kernels~\cite{Campbell:1997hg,Catani:1998nv,Braun-White:2022rtg}. Such
a correction will generate an extra term in Eqs.~\eqref{eq:GF-NLL}
which contains a product of three GFs (e.g. the splitting $q\to q gg$
will be proportional to $G_q\,G_g\,G_g$). At the same time one has to
consider the virtual one-loop corrections to the $1\to 2$ splitting
kernels~\cite{Sborlini:2013jba}, which will have the same GFs
structure as the NLL kernel~\eqref{eq:GF-NLL}. Finally, to avoid
double counting with the ${\cal O}(\alpha_s^2)$ iteration of the NLL
evolution, we must subtract the latter from the r.h.s. of
Eqs.~\eqref{eq:GF-NLL}.

The aforementioned calculation can be consistently performed in the
dimensional regularisation scheme in $d=4-2\epsilon$
dimensions. However, in order to exploit the flexibility of Monte
Carlo integration, we need to bring the integral equations into a form
that is manifestly finite so that we can take $\epsilon\to 0$ at the
integrand level.
In order to make the cancellation of $\epsilon$ divergences manifest,
we include a local subtraction term with the goal of regularising both
the $1\to 3$ real and $1\to 2$ virtual corrections.
The subtraction can be built directly from the $1\to 3$ real
corrections, by integrating them with the same $1\to 2$ GFs structure
present in the NLL kernel~\eqref{eq:GF-NLL}. In the reals, this effectively plays
the role of a virtual correction obtained by unitarity.
This procedure results in evolution equations that are manifestly
finite in four space time dimensions.
In the case of quark jets, the NNLL evolution equation takes the
form
\begin{align}\label{eq:Kvquark}
 G_q(x,t)= u\, \Delta_q(t) \,+&
                               \int_t^{t_0}d t'\int_{z_0}^{1-z_0}d z\, \,G_q(x\,z,t')
  \,G_g(x\,(1-z),t')\frac{\Delta_q(t)}{\Delta_q(t')}\,{\mathcal P}_{q}(z,\theta')\\
&+ {\mathbb K}_q^{\rm finite}[G_q,G_g] \,.\notag
\end{align}
In the above equation we defined the \textit{inclusive emission
  probability} ${\mathcal P}_{q}(z,\theta)$
as~\footnote{The precise form of $g(z)$ in the
  coefficient of $ {\mathcal B}^q_2(z)$ is not relevant at NNLL, but
  we keep it in for simplicity.}
\begin{align}\label{eq:quarkP}
{\mathcal P}_{q}(z,\theta)\equiv& \frac{2\, C_F}{1-z}\left(1 +
                                  \frac{\alpha_s(E^2g^2(z)\theta^2)}{2\pi}K^{(1)}\right)\notag\\
&+{\mathcal B}_1^q(z)+\frac{\alpha_s(E^2g^2(z)\theta^2)}{2\pi}\left({\mathcal B}^q_2(z)+{\mathcal B}_1^q(z) b_0\ln g^2(z)\right)\,,
\end{align}
where $\theta$ is the angle between the final state quark and gluon
(set by the evolution time $t'$).
The inclusive emission probability ${\mathcal P}_q(z)$ (and its
gluonic counterpart) will be also central to the construction of an
algorithmic solution to the NNLL problem, as we shall demonstrate in
forthcoming work~\cite{vanBeekveld:2023?}.
The quantity $K^{(1)}$ is the ratio
of the two-loop to the one-loop cusp anomalous dimension
\begin{equation}
K^{(1)}=\left(\frac{67}{18} - \frac{\pi^2}{6}\right)\,C_A -
\frac{10}{9}\,T_R\,n_f \equiv K^{(1),C_A}\,C_A + K^{(1),n_f}\,T_R\,n_f \,,
\end{equation}
and $b_0$ is the first coefficient of the QCD beta function
\begin{align}\label{eq:b0}
	b_0 = \frac{11}{6}\, C_A - \frac23 \,T_R \,n_f \equiv b_0^{(C_A)} \, C_A + b_0^{(n_f)}\, T_R \,n_f \,.
\end{align}
We have further decomposed the splitting function $P_{qq}(z)$ into its
soft part ($z\simeq 1$) and the hard-collinear left over, where the
latter is given by
\begin{equation}
{\mathcal B}_1^q(z) \equiv - C_F \,(1+z)\,.
\end{equation}
The term proportional to $b_0\,\ln^2 g(z)$ balances the $g(z)$
dependence of the argument of the overall coupling encoded in $d t'$
in Eq.~\eqref{eq:sud-quark}, such that the quantity
${\mathcal B}_2^q(z)$ is independent of the choice of $g(z)$ and
matches the definition and calculation given in
Ref.~\cite{Dasgupta:2021hbh}.
The definition of ${\mathcal P}_{q}(z,\theta)$ implicitly relies upon
a scheme to define the variables $z$ and $\theta$ while integrating
over a second emission (and adding the corresponding virtual
corrections). This scheme must be IRC safe and effectively
${\mathcal P}_{q}(z,\theta)$ can be thought of as a
next-to-leading-order correction to a $1\to 2$ collinear splitting.

The functional $ {\mathbb K}_q^{\rm finite}$ contains 4-dimensional
terms which can be handled efficiently with Monte Carlo
methods.\footnote{The application of this formalism to collinear
  unsafe quantities, such as fragmentation functions, entails
  considerable conceptual subtleties~\cite{vanBeekveld:2023?}.}
These terms satisfy a unitarity condition
$ \left.{\mathbb K}_q^{\rm finite}\right|_{u=1}=0$, they are at most NNLL
($\alpha_s^n L^{n-1}$) and contain structures of the $G_i\,G_j\,G_k$
type.
Although in the most general case they enter in NNLL resummations, for
event shapes and jet rates they only enter in the soft limit and are
responsible for correlated and clustering corrections in the language
of
Refs.~\cite{Banfi:2014sua,Banfi:2016zlc,Banfi:2018mcq,Anderle:2020mxj,Dasgupta:2022fim}. On
the other hand, for purely collinear problems, such as the dynamics of
small-$R$ jets~\cite{Dasgupta:2014yra}, ${\mathbb K}_q^{\rm finite}$
contributes as a whole.
The precise definition of ${\mathbb K}_q^{\rm finite}[G_q,G_g]$ goes
hand-in-hand with the definition of the function ${\mathcal B}_2^q$,
of which it specifies the scheme. Nevertheless, any physical prediction is
independent of such a scheme, and for any IRC safe observables the
scheme change can be performed directly in $d=4$ dimensions. We report
the expressions of such terms for quark and gluon jets in
Appendix~\ref{app:kernels}, while in the body of this article we focus
on the first line of Eq.~\eqref{eq:Kvquark} and its gluonic
counterpart.

The function $ {\mathcal B}^q_2(z)$ was calculated recently in
Ref.~\cite{Dasgupta:2021hbh} and will be reviewed in the next
section.
One can now define the Sudakov $\Delta_q(t)$ at NNLL from
Eq.~\eqref{eq:Kvquark}. We can set $u=1$ and obtain
\begin{align}\label{eq:sud-quark}
\ln \Delta_q(t) &= - \int_t^{t_0} d t' \int_{z_0}^{1-z_0}d z\, {\mathcal P}_{q}(z,\theta')\,.
\end{align}
The goal of this article is to define and calculate the analogous
quantity for gluon jets $ {\mathcal B}^g_2 (z)$, which paves the way
to obtain the Sudakov for the gluon fragmentation and the complete set
of integral equations which generalises jet calculus in the collinear
limit to NNLL.

Finally, the function $g(z)$ fixes the precise scale of the coupling
that becomes relevant at NNLL. An important constraint on $g(z)$ is
that in the soft limit ($z\simeq 1$) it is fixed
to~\cite{Catani:1990rr}
\begin{equation}\label{eq:limit}
  \lim_{z\to 1}\frac{g(z)}{1-z} = 1\,,
\end{equation}
so that the coupling is evaluated at the relative transverse momentum
of the branching.
Beyond the soft limit, the form of $g(z)$ and $ {\mathcal B}_2 (z)$
are linked and emerge from an ${\cal O}(\alpha_s^2)$ calculation. For
the sake of concreteness, in the rest of this article we take
\begin{equation}
  g(z)=1-z\,.
\end{equation}
This defines unambiguously the terms in $ {\mathcal B}_2 (z)$ that are
proportional to the QCD $\beta(\alpha_s)$ function at one loop
(i.e. $b_0$).

\section{$\mathcal{B}^q_2(z)$ in the quark case and the physical
  coupling scheme}
\label{sec:B2quark}
In this section we give a concise review of the results for quark jets
in Ref.~\cite{Dasgupta:2021hbh}. This helps elucidate the main
messages of the current work, in addition to showcasing some
differences to the case of gluon jets.
In the inclusive emission probability for quark
jets~\eqref{eq:quarkP}, the intensity of radiation in the soft limit
($z\simeq 1$) can be encoded in a physical scheme of the strong
coupling constant~\cite{Catani:1990rr}, in such a way that the
emission probability is defined as
\begin{align}\label{eq:probnllsoft}
	 \left. d t \,d z\, {\mathcal
  P}_{q}(z,\theta)\right|_{\rm soft}\simeq C_F \frac{\sd \theta^2}{\theta^2} \sd z\, \frac{2}{1-z} \,\frac{\alpha_s^{\text{CMW}}(E^2(1-z)^2\theta^2) }{2\pi} \ ,
\end{align}
where the r.h.s. amounts to using the first line of
Eq.~\eqref{eq:quarkP} and the CMW coupling is related to the
$\overline{\rm MS}$ one by the following relation
\begin{align}\label{eq:cmw}
	\alpha_s^{\text{CMW}}(\mu^2) = \alpha_s(\mu^2) \left(1 + \frac{\alpha_s(\mu^2)}{2\pi} K^{(1)} \right)\, .
\end{align}
The above scheme is customarily used in QCD resummations and is a
crucial analytical ingredient for NLL parton shower algorithms.

The extension of this picture to higher logarithmic accuracy is not
unique, and depends on the definition of the kinematic variables $z$
and $\theta$. While generalisations of Eq.~\eqref{eq:cmw} to three
loops (encoded in the coefficient $K^{(2)}$) in the soft limit
(relevant for double-logarithmic resummations) were obtained in
Refs.~\cite{Banfi:2018mcq,Catani:2019rvy}, much less is known about
its generalisation to the hard-collinear limit.
In resummation literature this information is encoded, at the
integrated level, in the \textit{observable-dependent} collinear
anomalous dimension commonly known as $B_2$, which nowadays has been
calculated for several
observables~\cite{Davies:1984hs,deFlorian:2001zd,deFlorian:2004mp,Banfi:2018mcq}.
When used in resummations of double logarithmic observables such as
event shapes or transverse momentum of a colour singlet, $B_2$ always
takes the form (see
e.g. Refs.~\cite{Davies:1984hs,deFlorian:2001zd,deFlorian:2004mp,Banfi:2018mcq})
\begin{align}
B_2^q  = - \gamma_q^{(2)} + b_0 X^q_v\ ,
\end{align}
where $\gamma_q^{(2)} $ is the endpoint of the singlet DGLAP splitting
kernel at two loops (e.g.~\cite{Ellis:1996mzs})
\begin{align}
	\gamma_q^{(2)} = C_F^2 \left(\frac38 - \frac{\pi^2}{2} + 6 \zeta_3\right) + C_F C_A \left(\frac{17}{24} + \frac{11 \pi^2}{18} -3 \zeta_3\right)- C_F T_R n_f \left(\frac16 + \frac{2\pi^2}{9}\right) \, ,
\end{align}
and the quantity $X^q_v$ is observable dependent.

Eq.~\eqref{eq:sud-quark} however suggests that we can encode this
observable dependence into the \textit{observable-independent}
inclusive emission probability ${\mathcal P}_{q}(z,\theta)$, which
encapsulates a differential anomalous dimension ${\cal B}^q_2(z)$. The
observable dependence of $B_2$ thus emerges entirely from the
integration over $z$, and specifically from integrating the
Sudakov~\eqref{eq:sud-quark} with the \textit{single-emission}
measurement function parameterised in terms of $z$ and $\theta$,
i.e.\footnote{We note that some definitions of the $B^q_2$ coefficient
  (and thus of $z$ and $\theta$) in the literature contain an extra
  contribution arising from single-logarithmic soft physics (see
  e.g. Ref.~\cite{Banfi:2018mcq}). These terms do not contribute to
  $B^q_2$ in our scheme for quark jets, and they emerge from the
  integration of the ${\mathbb K}_q^{\rm finite}[G_q,G_g]$
  contribution. We will return to this point when discussing the case
  of gluon jets.}
\begin{equation}\label{eq:Vztheta}
  \Theta(V(z,\theta)-v)\,.
\end{equation}

This offers a natural path to incorporate this information in Monte Carlo
parton showers, and it constitutes a crucial step towards the
development of NNLL algorithms.
Recently, Ref.~\cite{Dasgupta:2021hbh} presented a two loop
calculation of the quantity $ {\mathcal B}^q_2(z)$ in
Eq.~\eqref{eq:sud-quark} for the quark case, where the corresponding
$1\to 3$ splitting kernels are integrated by fixing $z$ and $\theta$
to the momentum fraction and angle of either the first emission,
i.e. the one at larger angles for the $C_F^2$ channel, or that of the
radiated pair (for the $C_F C_A$, $T_R n_f$ and $C_F\,(C_F-C_A/2)$
channels).\footnote{The definition of the radiated pair is ambiguous
in the $C_F\,(C_F-C_A/2)$ colour channel due to the symmetry of the
splitting kernel, however this ambiguity does not affect the form of
${\mathcal B}^q_2(z)$.}
Its expression is given in Appendix~\ref{app:B2q}.

The differential anomalous dimension $ {\mathcal B}^q_2(z)$ is
sufficient to derive $B_2$ for any rIRC safe global observable defined
on quark jets (see also the related discussions in Section 3 of
Ref.~\cite{Dasgupta:2021hbh}).
Specifically, for the definition of $z$ and $\theta$ adopted in
Ref.~\cite{Dasgupta:2021hbh} (and reported in Appendix~\ref{app:B2q})
one obtains
\begin{align}\label{eq:expform-quark}
  \int_0^1 dz\,{\cal B}^q_2(z) \equiv B_{2,\theta^2}^q  = - \gamma_q^{(2)} + b_0 X^q_{\theta^2}\ ,
\end{align}
where
\begin{align}\label{eq:xtheta-quark}
	X^q_{\theta^2} = C_F\left(\frac{2\pi^2}{3} -\frac{13}{2}\right) \, .
\end{align}
An analogous integration taking into account the observable
constraint~\eqref{eq:Vztheta} would produce the observable-specific
constant $X^q_v$. Explicit examples of this will be considered in
Sec.~\ref{sec:applications}.

\section{${\mathcal B}_2^g$ in the gluon case: definitions and
  computational strategy}\label{sect:defs}
To calculate $ {\mathcal B}^g_2(z)$, one needs an IRC safe definition
of $z$ and $\theta$ such that it projects the $1\to 3$ phase space
$\Phi_3$ onto the underlying $1\to 2$ kinematics $\Phi_2$. The
definition of $ {\mathcal B}^g_2(z)$ is then uniquely specified by a
kinematical map
\begin{equation}\label{eq:mapB2}
{\cal M}: \Phi_{3} \to \Phi_{2}\,,
\end{equation}
that provides a definition of the longitudinal momentum fraction $z$
and the angle $\theta$ of the first branching in the $\Phi_2$ phase space.
This also defines in a unique manner the inclusive emission
probability ${\mathcal P}(z,\theta)$, where one integrates over a
second emission while keeping the variables $z$ and $\theta$ fixed.
A property of the definition of $z$ and $\theta$ is that this
projection reproduces $K^{(1)}$ in the soft limit.
The generalisation of the calculation of the quark
case~\cite{Dasgupta:2021hbh} to the gluonic case is non-trivial and it
entails two conceptual subtleties:
\begin{enumerate}
\item In the case of quark jets it was possible to define $z$ and
  $\theta$ based on the colour structure of the triple collinear
  matrix elements. Specifically, independent (i.e. $C_F^2$) and
  correlated (i.e. $C_F\,C_A$, $C_F\,T_R$, and $C_F\,(C_F-C_A/2)$)
  channels are separated by their colour structure. This allows one to
  choose $z$ and $\theta$ such that in the abelian channel ($C_F^2$)
  these correspond to the kinematics of the first emission in an
  angular ordered picture, while in the remaining channels these
  variables correspond to the kinematics of the \textit{parent}
  emitter of the final state pair of real partons (see
  e.g. Fig.~\ref{fig:a-splits}).
  In the gluonic case, and specifically for the $C_A^2$ piece, this
  correspondence between independent and correlated emissions and
  different colour channels is no longer valid, in that independent
  and correlated radiation are mixed under the same colour structure.
  As a result, we need to formulate an IRC safe procedure to define
  $z$ and $\theta$ in all gluonic channels.
\item A second subtlety concerns the flavour structure and the
  divergences in the gluonic case, which is more involved than in the
  quark case as reflected in the evolution equation for the gluonic
  GF~\eqref{eq:GF-NLL}. Unlike in the quark case, the gluon GF encodes
  two different splitting kernels already at NLL. Of these,
  $p_{gg}(z)$ is such that in an iterated $1\to 2$ splitting
  $g\to g_a (z) g_b(1-z)$ followed by, e.g., $g_b\to q\bar{q}$ the
  gluon $g_a$ which does not branch can still bring a soft divergence
  (this time at $z\simeq 0$). This behaviour is absent in the quark
  case (as $z\to 0$ would correspond to a soft quark configuration),
  and ensuring the cancellation of all IRC divergences in the gluon
  case requires a definition of the underlying $1\to 2$ branching that
  is safe in this limit.
\end{enumerate}
We give below a definition of $z$ and $\theta$ that addresses the
above features and can be used to derive $ {\mathcal B}^g_2(z)$.
The procedure follows a strategy inspired by the mMDT/Soft-Drop
$(\beta=0)$ algorithm~\cite{Dasgupta:2013ihk,Larkoski:2014wba} which
was designed to identify hard-collinear splittings in a jet. It can be
summarised in the following steps:
\begin{itemize}
\item For each colour channel, decompose the phase space into angular sectors, such that each sector contains exactly one collinear singularity. Evidently, the number of required sectors matches the number of singular collinear configurations in each colour channel.
\item For each sector, we cluster the two partons, say $i$ and $j$, which give rise to the collinear singularity as $\theta_{ij} \to 0$.
\item In each sector the angle $\theta$ is identified with the
  angle between the clustered pair $(ij)$ and the remaining
  parton $k$. Similarly, the longitudinal momentum fraction $z$
  coincides with that of parton $k$, i.e. $z \equiv z_k$.
  We then symmetrise the result by adding the permutation $z\to 1-z$
  and multiply by $1/2!$ if the Born $1\to 2$ splitting to which we
  are projecting involves identical particles.
  The procedure follows the angular-ordered pattern of the
  generating functional evolution equation.
\item In the region of phase space where $z_k \to 0$, the definition
  of the angle $\theta$ becomes unsafe. Therefore, when $z_k$ is below
  a certain threshold, say $z_{\text{cut}} \ll 1$, we discard the soft
  branch containing $z_k$ and fix the angle
  $\theta \equiv \theta_{ij}$. In addition the longitudinal momentum
  is chosen as $z \equiv z_i/(z_i+z_j)$ which must necessarily pass
  the $z_{\text{cut}}$ condition in order to define a hard-collinear
  splitting.
\item Finally, we explicitly take the limit $z_{\rm cut}\to 0$.   
\end{itemize}
In practice we consider the fragmentation of a gluon jet defined as
one hemisphere in the two-jet limit of the $H\to gg$ decay in the
heavy-top-mass limit. The corresponding NNLL resummed
doubly-differential distribution in $z$ and $\theta$ is given in
Eq.~\eqref{eq:theta-z-dist} of Appendix~\ref{app:SDquark}.
The extraction of $ {\mathcal B}^g_2(z)$ then can proceed by
simply relating the second order calculation for
$\frac{\theta^2}{\sigma_0}\frac{d^2\sigma}{d\theta^2d z} $ to the
second order expansion of Eq.~\eqref{eq:theta-z-dist}.
In general our procedure is not exactly the same as mMDT~\footnote{An
  exception is given by the $C_A^2$ channel, where our procedure is
  equivalent to the mMDT algorithm.} since it does
not correspond to the exact C/A declustering sequence. Yet, our
clustering sequence is sufficient to capture all the divergent
structure and obtain an IRC safe effective emission probability.

In contrast to the quark case, in which the integral of
${\cal B}^q_2(z)$ yields the simple form given in
Eq.~\eqref{eq:expform-quark}, the richer singularity structure present
in the gluon case (specifically in the $C_A^2$ channel) and reflected
in the emergence of multiple singular sectors as outlined above will
lead to the general form
\begin{equation}
  \label{eq:expform-gluon}
\int_0^1 {\cal B}^g_2(z) \equiv B_{2,\theta^2}^g   = -\gamma_g^{(2)}+ b_0 X^g_{\theta^2}+ \mathcal{F}_{\text{clust.}}^{C_A^2}\,,
\end{equation}
where~\cite{Ellis:1996mzs}
\begin{equation}
\label{eq:b2singlet}
-\gamma_g^{(2)} = \frac{4}{3} C_A T_R n_f+C_F T_R n_f-C_A^2 \left(\frac{8}{3}+3\zeta_3 \right),
\end{equation}
and $X^g_{\theta^2}$ is a constant that we shall determine in
Sec.~\ref{sect:b2extraction}.
Finally, the extra term $\mathcal{F}_{\text{clust.}}^{C_A^2}$ emerges
from the non-trivial sectorisation that becomes necessary in the
$C_A^2$ channel (cf. Sec.~\ref{sec:B2ca2}).

In the next sections we discuss how to use the above algorithm to
define and calculate the inclusive emission probability
$\mathcal{P}_g(z,\theta)$.
In particular, in order to calculate $\mathcal{P}_g(z,\theta)$ we
start by computing the doubly-differential distribution
$\theta^2 d^2\sigma/d\theta^2dz$, and then in
Sec.~\ref{sect:b2extraction} we will extract
$\mathcal{P}_g(z,\theta)$.
We will discuss separately the contribution of the one-loop correction
to the collinear $1 \to 2$ splitting kernel and of the tree-level
$1 \to 3$ splitting kernels.

\section{Virtual corrections to $1\to 2$ collinear splitting}
We start by discussing the ${\cal O}(\alpha_s^2)$ real-virtual
corrections to the $\theta^2 d^2\sigma/d\theta^2dz$ distribution.
We parameterise them as follows:
\begin{align}\label{eq:RV}
  \left(\frac{\theta^2}{\sigma_0} \frac{\sd^2\sigma_{\mathcal{V}}^{(2)}}{\sd \theta^2 \, \sd z}\right)  = \left(\frac{\alpha_s}{2\pi}\right)^2 \mathcal{V}(\theta^2,z,\epsilon) \ ,
\end{align}
where, in this case, the kinematical variables $(\theta,z)$ are
uniquely defined as those of a $1 \to 2$ collinear splitting.
Our results will be expressed in terms of a renormalised
$\overline{\text{MS}}$ coupling related to the bare strong coupling by
\begin{equation}
	\label{eq:renorm}
	\mu^{2\epsilon} \alpha_s = S^{-1}_\epsilon\mu_R^{2\epsilon} \alpha_s(\mu_R^2) \left(1-\frac{b_0}{\epsilon}\frac{\alpha_s(\mu_R)}{2\pi}+\mathcal{O}(\alpha_s^2)\right)\,,
\end{equation}
where
\begin{equation}
  \label{eq:msbar}
  S_\epsilon = (4\pi)^\epsilon e^{-\epsilon \gamma_E}\,.
\end{equation}
Finally, we choose $\mu_R =E$ (the energy of the hard parton
initiating the fragmentation) to present the results, and define
$\alpha_s \equiv \alpha_s (E^2)$.
There are three kinds of one-loop corrections that are involved in the
calculation, and they can be separated according to their origin. The
first is the one-loop UV counter-term that is related to the
renormalisation of the bare coupling~\eqref{eq:renorm}, and reads:
\begin{equation}
  \label{eq:onelooprenorm}
  \mathcal{V}_{\text{renorm.}}(\theta^2,z,\epsilon) = - \frac{b_0}{\epsilon} \left(z(1-z)\theta \right)^{-2\epsilon} \left(C_A\, p_{gg}(z) + T_R n_f\, p_{qg} (z,\epsilon) \right)\,.  \\
\end{equation}
The second is the one-loop correction to the $1 \to 2$ splitting
function itself. These corrections are taken from
Ref.~\cite{Sborlini:2013jba}, in the CDR scheme.\footnote{In the
  $T_R^2 n_f^2$ channel, we noticed a typographical error in
  Eq.~(5.19) of the journal version of Ref.~\cite{Sborlini:2013jba},
  which we fix to recover the divergent structure predicted by
  Catani's one-loop formula~\cite{Catani:1998bh}. Specifically, one
  should replace $-3$ with $+3$ in the denominator of the first term
  on the r.h.s. of Eq.~(5.19) in the journal version of
  Ref.~\cite{Sborlini:2013jba}.}
Explicitly, the one-loop correction to the collinear $g \to q \bar{q}$
splitting is given by
\begin{multline}\label{eq:gqq1loop}
	\mathcal{V}^{(1)}_{g \to q \bar{q}}(\theta^2,z,\epsilon) = (z(1-z))^{-3\ep} \theta^{-4\ep}\, T_R n_f\, p_{qg}(z,\ep) \left(\frac{1}{\ep^2}- \frac{2\pi^2}{3}\right) \times \\
	\times \bigg[ T_R n_f \, f_{T_R n_f}^{g \to q \bar{q}} 
	+ C_F \, f_{C_F}^{g \to q \bar{q}} 
	+ C_A \, f_{C_A}^{g \to q \bar{q}} \bigg] \ ,
\end{multline}
where
\begin{align}
	f_{T_R n_f}^{g \to q \bar{q}} &= -\frac{4\ep}{3} - \frac{20 \ep^2}{9}  + \mathcal{O}\left(\ep^3\right) \ , \\
	f_{C_F} ^{g \to q \bar{q}}&= -2-3\ep-8\ep^2 + \mathcal{O}\left(\ep^3\right) \ , \\
	f_{C_A}^{g \to q \bar{q}} &= 1 + \frac{11\ep}{3}+ \frac{76 \ep^2}{9}+\ep \ln(z(1-z)) + \ep^2\, \text{Li}_2\left(\frac{z-1}{z}\right)+ \ep^2 \,\text{Li}_2\left(\frac{z}{z-1}\right)+ \mathcal{O}(\ep^3) \,.
\end{align}
For the one-loop correction to the collinear $g \to gg$ splitting we
have
\begin{align}\label{eq:gggo1loop}
	\mathcal{V}^{(1)}_{g \to gg}(\theta^2,z,\epsilon) =
  (z(1-z))^{-3\ep} \theta^{-4\ep} C^2_A \, p_{gg}(z,\ep)
  \left(\frac{1}{\ep^2}- \frac{2 \pi^2}{3}\right) \, f_{C_A}^{g \to
  gg} + \frac16 C_A^2 - \frac13 C_A T_R n_f \ ,
\end{align}
where
\begin{align}
	f_{C_A}^{g \to gg}  = -1 + \ep \ln(z(1-z)) + \ep^2\, \text{Li}_2\left(\frac{z-1}{z}\right)+ \ep^2\, \text{Li}_2\left(\frac{z}{z-1}\right)  + \mathcal{O}\left(\ep^3\right)\,.
\end{align}
We finally note that if we were to compute the full
$\mathcal{O}(\alpha_s^2)$ result for a given observable then the
one-loop virtual corrections to the Born process would also have to be
included. In practice, for the double-differential calculation at
second order that we carry out here only the product of such one-loop
virtual corrections with the single collinear splitting function is needed.
For the $H\to gg$ process in the heavy-top effective theory considered
in this article it reads (with $\mu_R=E$)\footnote{\label{footnote:effective-coupling}Note that the finite part of Eq.~\eqref{eq:oneloopborn} does not include the contribution from the finite term of the one-loop matching coefficient in the $Hgg$ effective coupling. This choice is not relevant in the extraction of $\mathcal{B}_2(z)$, as the process dependence cancels.}
\begin{align}\label{eq:oneloopborn}
\mathcal{V}^{(1)\,{\rm Born}}_{H \to gg}(\theta^2,z,\epsilon) &=
\left(C_A\,p_{gg}(z)+T_R
  \,n_f\,p_{qg}(z,\epsilon)\right)\,\left(1-\frac{\pi^2}{12}\epsilon^2\right)(z^2(1-z)^2\theta^2)^{-\epsilon}\notag\\
  & \times \left(\left(-\frac{4^{-\epsilon}}{\epsilon^2} -\frac{11}{6\epsilon} + \frac{7}{12}\pi^2\right)\,C_A+\frac{2}{3\epsilon}T_R\,n_f\right)\,.
\end{align}
In the above expression, the process dependence is embodied in the
${\cal O}(\epsilon^0)$ terms of the above equation, and it cancels in
the extraction of ${\cal B}^g_2(z)$ when taking the difference between
our second-order calculation and the expansion of the resummation
formula given in Eq.~\eqref{eq:theta-z-dist}.
The function $\mathcal{V}(\theta^2,z,\epsilon)$ in Eq.~\eqref{eq:RV}
then reads
\begin{equation}
\mathcal{V}(\theta^2,z,\epsilon) = \mathcal{V}_{\text{renorm.}}(\theta^2,z,\epsilon)+\mathcal{V}^{(1)}_{g \to q \bar{q}}(\theta^2,z,\epsilon)+\mathcal{V}^{(1)}_{g \to gg}(\theta^2,z,\epsilon) +\mathcal{V}^{(1)\,{\rm Born}}_{H \to gg}(\theta^2,z,\epsilon) \,.
\end{equation}

\section{Real $1 \to 3$ collinear splitting}
\label{sect:1to3}
In this section we discuss the integration of the triple-collinear
splitting functions over the three-body phase space, while keeping
fixed $z$ and $\theta$ as discussed in Sec.~\ref{sect:defs}. We start
by introducing the spin-averaged triple-collinear splitting functions
relevant for gluon fragmentation. By considering the spin-averaged
case, we implicitly focus on observables that are insensitive to spin
correlations.%
\footnote{Note that our formalism is also applicable to spin-sensitive observables, in which case one would need to use the polarised $1\to 2$ and $1\to 3$ splitting kernels as ingredients in the calculations (including that of $\mathcal{B}^f_2(z)$). The $1\to3$ splitting kernels used here contain partial spin information, which is sufficient to obtain LL accuracy for spin sensitive observables.}
For an initial gluon there are three distinct splitting
functions to consider organised by colour structure. In the notation
of Ref.~\cite{Catani:1998nv} these are:
\begin{itemize}
	\item $\langle \hat{P}_{g_1 g_2 g_3}  \rangle$ corresponding to a $1 \to 3$ gluon splitting into three gluons.
	\item $\langle \hat{P}^{\textrm{(ab)}}_{g_1 q_2 \bar{q}_3}
          \rangle$ corresponding to the abelian channel, $C_F T_R\,n_f$, of
          a gluon splitting into a $q\bar{q}$ pair, with subsequent
          emission of a gluon.
	\item $\langle \hat{P}^{\textrm{(nab)}}_{g_1 q_2 \bar{q}_3}
          \rangle$ corresponding to the non-abelian channel,
          $C_A T_R\,n_f$, of a gluon splitting into two gluons, one of
          which branches further into a $q\bar{q}$ pair.
\end{itemize}
The splitting functions in Ref.~\cite{Catani:1998nv} are expressed in
terms of invariants, $s_{ij}$, and energy fractions $z_i$. In the
relevant collinear limit, we have that
$s_{ij} = (p_i+p_j)^2 \simeq E^2 z_i z_j \theta_{ij}^2$ where $E$
denotes the energy of the initial gluon. The triple-collinear phase
space in $d=4-2\epsilon$ dimensions may be expressed in the form
\cite{Gehrmann-DeRidder:1997fom}
\begin{align}\label{eq:psnormal}
	\sd \Phi_3 = \frac{1}{\pi}  \frac{E^{4-4\epsilon}}{(4\pi)^{4-2\epsilon} \Gamma(1-2\epsilon)} \sd z_2 \sd z_3 \sd \theta_{13}^2 \sd  \theta_{23}^2 \sd \theta_{12}^2 (z_1z_2 z_3)^{1-2\epsilon} \Delta^{-1/2-\epsilon} \,\Theta(\Delta) \ \ ,
\end{align}
where the Gram determinant is given by
\begin{align}
	\Delta = 4 \theta_{ik}^2 \theta_{jk}^2 - \left(\theta_{ij}^2 - \theta_{ik}^2- \theta_{jk}^2\right)^2 ,\quad i\neq j \neq k\,,
\end{align}
and
\begin{equation}
\sum_{i=1}^3z_i=1\,.
\end{equation}
As explained before, we are interested in the double-differential
distribution in $(z,\theta)$, which is defined according to the
procedure of Sec.~\ref{sect:defs} and will be illustrated in the
following:
\begin{itemize}
\item $C_A T_R\,n_f$: This colour channel exhibits a simple collinear
  structure because the sole collinear singularity appears as
  $\theta_{23} \to 0$, see
  Fig.~\ref{fig:triple-collinear-nonab}. Therefore, we fix the energy
  fraction of the gluon to be $z_1 = z$ and the angle
  $\theta = \theta_g$ shown in
  Fig.~\ref{fig:triple-collinear-nonab}. We map the set
  $(z_1,z_2,z_3)$ into an independent set $(z,z_p)$ defined as shown
  in the figure. Instead when $z$ falls below $z_{\text{cut}}$, the angle
  will be defined as $\theta = \theta_{23}$ and the longitudinal
  momentum defined as $z = z_2$ as shown in Fig.~\ref{fig:catrfail}.
  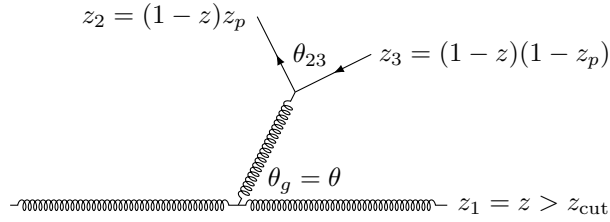
\begin{figure}[h!]
	\centering
	\begin{tikzpicture}[scale=2.5] 
		
		\coordinate (bq1) at (0,0);
		\coordinate (bq2) at (1.2,0); 
		\coordinate (bq3) at (2.3,0);
		
		\coordinate (bg1) at (1.2,0);
		\coordinate (eg1) at (1.5,0.6);
		\coordinate (bg2) at (1.5,0.6);
		\coordinate (eg2) at (1.3,1.0);
		\coordinate (bg3) at (1.5,0.6);
		\coordinate (eg3) at (1.9,0.8);
		
		\draw [gluon] (bq1) -- (bq2);
		\draw [gluon] (bg1) -- (eg1) ;
		\node at (1.55,0.13) {\small $\theta_g  = \theta$};
		\draw [quark] (bg2) -- (eg2) node [pos=1,left] {\small $z_2=(1-z)z_p$};
		\node at (1.58,0.8) {\small $\theta_{23}$};
		\draw [quark] (eg3) -- (bg3);
		\node at (2.55,0.8) {\small $z_3=(1-z)(1-z_p)$};
		\draw [gluon] (bq2) -- (bq3) node
		[pos=1,right] {\small $z_1 = z > z_{\text{cut}}$} ;

	\end{tikzpicture}
	\caption{The Feynman diagram representing gluon emission followed by its subsequent decay to a $q\bar{q}$ pair.}\label{fig:triple-collinear-nonab}
\end{figure}
      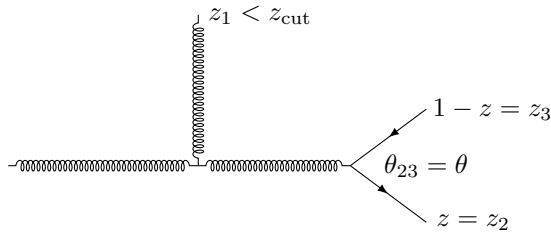
\begin{figure}[h]
	\centering
	\begin{tikzpicture}[scale=2.5] 
		
		\coordinate (bg1) at (0,0);
		\coordinate (bg2) at (1.0,0); 
		\coordinate (eg2) at (1.8,0);
		\coordinate (eg3) at (1.0,0.8);
		
		\coordinate (eq1) at (2.2,-0.3);
		\coordinate (bqb1) at (2.2,0.3);

		\draw [gluon] (bg1) -- (bg2);
		\draw [gluon] (bg2) -- (eg3)  node
		[pos=1,right] {\small $z_1 < z_{\text{cut}}$};
		\draw [gluon] (bg2) -- (eg2) ;
		
		\draw [quark] (eg2) -- (eq1)  node
		[pos=1,right] {\small $z=z_2$};
		\draw [quark] (bqb1) -- (eg2) ;
		
		\node at (2.2,0) {\small $\theta_{23} = \theta$} ;
		\node at (2.55,0.3) {\small $1-z =z _3$};
		
	\end{tikzpicture}
	\caption{The Feynman diagram representing gluon emission
          followed by its subsequent decay to a $q\bar{q}$ pair, where
          $g_1$ goes below the energy threshold and is allowed to fly
          off at wide angles.}
	\label{fig:catrfail}
      \end{figure}

    \item $C_F T_R\,n_f$: This colour channel exhibits two collinear
      singularities as $\theta_{13} \to 0$ and $\theta_{12} \to 0$,
      see Fig.~\ref{fig:triple-collinear-ab}. Therefore, we partition
      the phase space into two sectors $\theta_{13} < \theta_{12}$ and
      $\theta_{12} < \theta_{13}$. In the former, we fix the quark
      energy $z_2 = 1-z$ and the angle $\theta = \theta_{2,13}$. In
      the latter we fix the anti-quark energy $z_3 = z$ and the angle
      $\theta = \theta_{3,12}$. The configurations in which
      $z < z_{\text{cut}} $ ($z > 1- z_{\text{cut}} $) in sector 1 (2)
      correspond to power-suppressed contributions in $z_{\text{cut}}$
      (due to a soft quark) and are therefore neglected.
          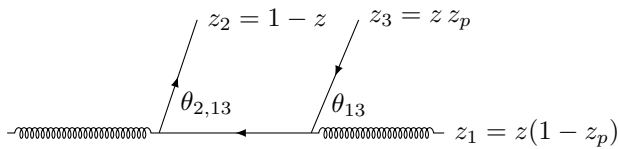
\begin{figure}[h!]
	\centering
	\begin{tikzpicture}[scale=2.5] 
		
		\coordinate (bg1) at (0.2,0);
		\coordinate (bg2) at (1,0); 
		\coordinate (bg3) at (1.8,0);
		\coordinate (eg3) at (2.5,0);

		\coordinate (bq1) at (1.0,0);
		\coordinate (eq1) at (1.2,0.6);
		\coordinate (bq2) at (1.8,0);
		\coordinate (eq2) at (2.05,0.6);
		
		\draw [gluon] (bg1) -- (bg2);
		\draw [quark] (bq1) -- (eq1) node [pos=1,right] {\small $z_2=1-z$} ;
		\node at (1.25,0.15) {\small $\theta_{2,13}$};
		\draw [quark] (eq2) -- (bq2) node [pos=-0.0,right] {\small $z_3 =z \, z_p$};
		\node at (2.0,0.15) {\small $\theta_{13}$};
		\draw [quark] (bg3) -- (bg2);
		\draw [gluon] (bg3) -- (eg3) node
		[pos=1,right] {\small $z_1 = z (1-z_p)$};

	\end{tikzpicture}
	\caption{The Feynman diagram representing the $q\bar{q}$ emission, and the energy fraction parameterisation is suitable for the region $\theta_{13} < \theta_{12}$.}
	\label{fig:triple-collinear-ab}
      \end{figure}
      
    \item $C_A^2$: This channel is the most complicated since there is
      a collinear pole as any angle $\theta_{ij} \to 0$. Therefore, we
      partition the phase space into three distinct sectors defined by
      the smallest angle in each. For example, in the sector
      $\text{min}\{\theta_{ij}\} = \theta_{12}$ we fix the angle
      between the parent of the clustered pair and the remaining
      gluon, i.e. $\theta =\theta_{12,3}$ as shown in
      Fig.~\ref{fig:triple-collinear-ca2}. The above sectoring removes
      a $1/3$ combinatorial factor, and in order to remove the
      remaining twofold symmetry in the $g\to ggg$ splitting function
      we require that $z_p \geq 1/2$ in the parameterisation of
      Fig.~\ref{fig:triple-collinear-ca2}.\footnote{We observe that
        any choice of the angle that is kept fixed to $\theta$ which
        is equivalent in the singular limits will produce the same
        result for ${\cal B}^g_2(z)$, e.g. one could also choose to fix
        $\theta=\theta_{13}$, and would only differ by further
        subleading (N$^3$LL) corrections.} The energy fraction being fixed is
      $z_3 = z$. When $z$ falls below $z_{\text{cut}}$, then
      $\theta = \theta_{12}$ and $z=z_2$.
          \begin{figure}[h]
	\centering
	\begin{tikzpicture}[scale=2.5] 
		
		\coordinate (bgg1) at (0,0);
		\coordinate (bgg2) at (1.2,0); 
		\coordinate (egg3) at (2.4,0);
		
		\coordinate (bg1) at (1.2,0);
		\coordinate (eg1) at (1.5,0.6);
		\coordinate (bg2) at (1.5,0.6);
		\coordinate (eg2) at (1.3,1.0);
		\coordinate (bg3) at (1.5,0.6);
		\coordinate (eg3) at (1.9,0.8);
		
		\draw [gluon] (bgg1) -- (bgg2);
		\draw [gluon] (bgg2) -- (egg3) node
		[pos=1,right] {\small $z_3 = z$} ;
		\draw [gluon] (bg1) -- (eg1) ;
		\node at (1.5,0.15) {\small $\theta_{12,3}$};
		\draw [gluon] (bg2) -- (eg2) node [pos=1,left] {\small $z_1=(1-z)z_p$};
		\node at (1.58,0.8) {\small $\theta_{12}$};
		\draw [gluon] (bg3) -- (eg3) node [pos=1,right] {\small $z_2=(1-z)(1-z_p)$};

	\end{tikzpicture}
	\caption{The Feynman diagram representing gluon emission
          followed by its subsequent decay to a $gg$ pair, where the
          energy fraction parameterisation suitable for the angular
          region $\text{min}\{\theta_{ij}\} = \theta_{12}$ is
          shown.}\label{fig:triple-collinear-ca2}
      \end{figure}
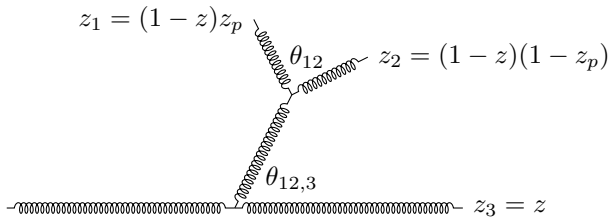
\end{itemize}

We integrate the splitting functions over the three-body phase space
in $d=4-2\epsilon$ dimensions to obtain real emission contributions.
The integrals we carry out are generically of the form
\begin{equation}
  \frac{\theta^2}{\sigma_0}\frac{\sd^2\sigma}{\sd z \, \sd \theta^2} = \int \text{d}\Phi_3(z_i,\theta_{ij}) \frac{\left(8\pi \alpha_s \mu^{2\epsilon}\right)^2}{s^2_{123}}\,  \langle \hat{P} \rangle \,\theta^2 \, \delta \left ( \theta^2 -\theta^2 \left(z_i,\theta_{ij} \right)\right) \delta(z-z(z_i))\, \Theta_{\text{cut}}(\theta_{ij})\,,
\end{equation}
where, $s_{123} = s_{12} + s_{13} + s_{23}$ and $\Theta_{\text{cut}}(\theta_{ij})$ denotes the angular cuts due
to sectoring that we described above.
We evaluate the above integrals by expanding out the singular
structure as a Laurent series in distributions, and then handle the
remaining integrations with {\tt Mathematica}.

\subsection{The $C_F  T_R \,n_f$ channel}
We start with the $C_F T_R \,n_f$ channel. The splitting function to be
integrated is the unpolarised quantity
$\langle \hat{P} _{g_1 q_2 \bar{q}_3} \rangle$ from
Ref.~\cite{Catani:1998nv}. The approach to the calculation of the
corresponding phase-space integrals, as well as other computational
details, are discussed in the corresponding calculation for the quark
fragmentation given in Refs.~\cite{Anderle:2020mxj,Dasgupta:2021hbh}.

\begin{figure}
	\centering
       \includegraphics[width=0.7 \textwidth]{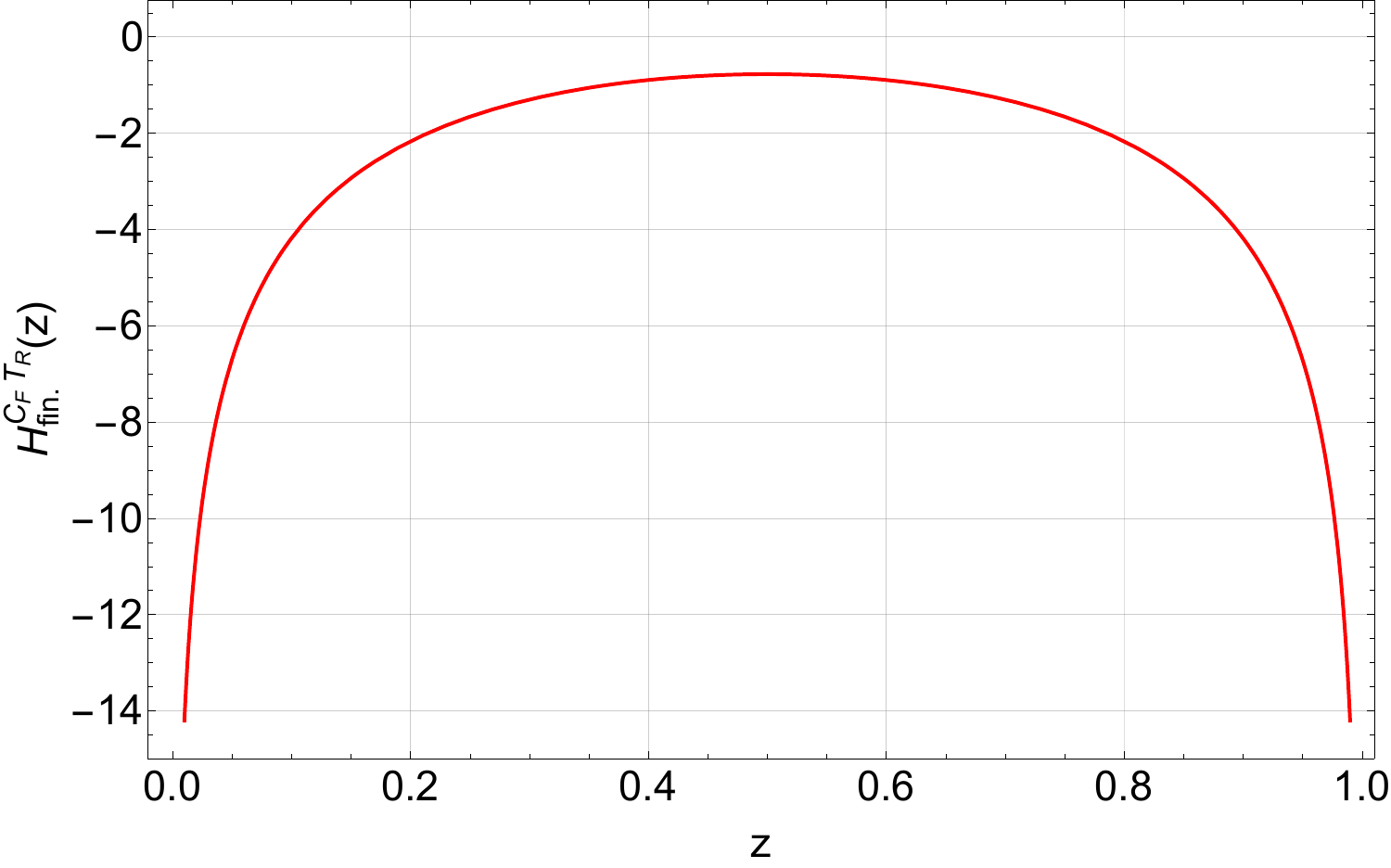}
       \caption{The function $H^{C_F T_R}_{\text{fin.}}
         (z)$.}
	\label{fig:hfincftf}
\end{figure}

Following our definition of $z$ and $\theta$, we consider two sectors:
$\theta_{13} < \theta_{12}$, which contains the collinear singularity
along the antiquark, and $\theta_{12} < \theta_{13}$, which contains
the collinear singularity along the quark. In the first sector we
parameterise the momenta as if the gluon is emitted from the antiquark
so that $z_2 =(1-z)$, $z_3 = z z_p$ and $z_1=z(1-z_p)$ as shown in
Fig.~\ref{fig:triple-collinear-ab}. In the second sector we instead
parameterise the energy fractions as if the gluon is emitted from the
quark, i.e. $z_3=z$, $z_1= (1-z_p)(1-z)$ and $z_2 = z_p (1-z)$. We
also fix the angle $\theta \equiv \theta_{2,13}$ in the first sector,
and $\theta \equiv \theta_{3,12}$ in the second sector. We can just
perform the calculation in the first sector and obtain the answer in
the second sector by sending $z \to 1-z$.
We report below the real emission contribution for the sector
$\theta_{13} < \theta_{12}$ which can be expressed as:
\begin{multline}\label{eq:cftfbasic}
\left(\frac{\theta^2}{\sigma_0} \frac{\sd^2\sigma_\mathcal{R}^{(2)}}{\sd\theta^2 \, \sd z}\right)^{C_F T_R, \mathrm{sec. 1}} = C_F  T_R n_f \left( \frac{\alpha_s}{2\pi} \right)^2   \bigg( \frac{H_{\text{soft-coll.}}^{\mathrm{sec. 1}} (\theta^2,z,\epsilon)}{ \epsilon^2}
+ \frac{H_{\mathrm{coll.}}^{\mathrm{sec. 1}}(\theta^2,z,\epsilon)}{\epsilon} \\ +\frac{H_{\mathrm{soft}}^{\mathrm{sec. 1}}(\theta^2,z,\epsilon)}{\epsilon} + H^{\mathrm{sec. 1}}_{\text{fin.}}(z) \bigg)\,,
\end{multline}
where the functions are labeled according to the origin of the
singular behaviour and $H^{\mathrm{sec. 1}}_{\text{fin.}}(z)$ is a
finite function that we cast as a 1-fold integral over $z_p$. As the
gluon is emitted from the (anti)-quark, the singular structure is
identical to that for the $C_F^2$ quark jet case, and reported in
Refs.~\cite{Anderle:2020mxj,Dasgupta:2021hbh}, {\em viz.}
\begin{equation}
\label{eq:cftfsing}
\begin{split}
H^{\mathrm{sec. 1}}_{\text{soft-coll.}}(\theta^2,z,\epsilon) &= z^{-4\epsilon} (1-z)^{-2\epsilon} \theta^{-4\epsilon}\,  p_{qg}(z,\epsilon) 
\left(1-\frac{\pi^2}{6}\epsilon^2+\mathcal{O}\left(\epsilon^3\right)\right) \ ,\\
H_{\text{coll.}}^{\mathrm{sec. 1}}(\theta^2,z,\epsilon) &= z^{-4\epsilon} (1-z)^{-2\epsilon} \theta^{-4\epsilon}\, p_{qg}(z,\epsilon)
\left(\frac{3}{2}+\frac{13}{2}\epsilon-\frac{2\pi^2}{3}\epsilon+\mathcal{O}\left(\epsilon^2\right)\right) \ ,\\
H_{\text{soft}}^{\mathrm{sec. 1}}(\theta^2,z,\epsilon) &= 0 \ .
\end{split}
\end{equation}
The result in the second sector can be obtained quite easily from the
first by sending $z \to 1-z$. In particular,
\begin{align}
	H^{\mathrm{sec. 2}}_{\text{fin.}}(z) = H^{\mathrm{sec. 1}}_{\text{fin.}}(1-z) \ .
\end{align}
We plot the total finite function,
$H^{C_F T_R}_{\text{fin.}} \equiv H^{\mathrm{sec. 1}}_{\text{fin.}} +
H^{\mathrm{sec. 2}}_{\text{fin.}}$ in Fig.~\ref{fig:hfincftf}. The
divergent behavior of $H^{C_F T_R}_{\text{fin.}} (z)$ as $z \to 0$ or
$z \to 1$ is only logarithmic, and thus it is integrable over
$z\in [0,1]$.  The integral read
\begin{align}
	\int_0^1 \sd z \, H^{C_F T_R}_{\text{fin.}} (z) = -3 \ .
\end{align}

\subsection{The $C_A T_R \,n_f$ channel} 
We start with the configuration depicted in
Fig.~\ref{fig:triple-collinear-nonab}.
For this channel, a convenient parametrisation of the three
body phase space can be obtained in terms of web variables as
discussed in Ref.~\cite{Dasgupta:2021hbh}.
The only collinear singularity in this channel emerges from the
collinear $g \to q\bar{q}$ splitting, i.e.~when $\theta_{23} \to
0$. Therefore, following our procedure outlined in
Sec.~\ref{sect:defs}, there is a single sector and thus we can
integrate inclusively over the angular phase space.
In this channel we need to consider the two situations depicted in
Figs.~\ref{fig:triple-collinear-nonab} and~\ref{fig:catrfail}, in
which the branching with a larger angle either passes or does not pass
the $z_{\text{cut}}$ threshold.
In the first case, the longitudinal momentum kept fixed is $z_1 = z$
(with $1-z_{\text{cut}} > z > z_{\text{cut}}$).
We fix the angle $\theta$ to be that of the parent of the $(23)$ pair,
$\theta=\theta_{23,1}=\theta_g$, which reads
 \begin{align}\label{eq:thetagdef}
\theta_{g}^2 = \frac{z_2}{z_2 + z_3} \theta^2_{12} + \frac{z_3}{z_2 + z_3} \theta^2_{13} - \frac{z_2 z_3}{(z_2+z_3)^2} \theta^2_{23} \ .
 \end{align}
 Finally, we symmetrise in $z\leftrightarrow 1-z$ and multiply by
 $1/2!$ to obtain:
\begin{align}\label{eq:catrfinalreal}
	\left(\frac{\theta^2}{\sigma_0} \frac{\sd^2\sigma_{\mathcal{R}}^{(2)}}{\sd\theta^2 \, \sd z}\right)_{z_1 > z_\text{cut}}^{C_A T_R}  &= \frac{1}{2!} C_A T_R n_f \left(\frac{\alpha_s}{2\pi}\right)^2  \, z^{-3\ep} (1-z)^{-3\ep}\, \theta^{-4\epsilon} \bigg[p_{gg}(z) \left(-\frac{8}{3\ep} 
	- \frac{40}{9}\right)\notag\\
	&\hspace{-1cm}-\frac43 (1+z)\, \ln z -\frac43 (2-z)\, \ln (1-z)
	+\frac{26}{9} \left(z^2+(1-z)^2-\frac{1}{z(1-z)}\right)
        +\frac{10}{3}  \bigg]\notag\\
        &\hspace{-1cm}\times\Theta(z - z_{\text{cut}})\,\Theta(1 - z_{\text{cut}}-z)\ .
\end{align}
In the second case (see Fig.~\ref{fig:catrfail}), the first gluon
fails the $z_{\text{cut}}$ condition. Accordingly, $z$ and $\theta$
are now fixed by the kinematics of the hard $g \to q\bar{q}$ splitting.
We note that, in contrast to the $C_F T_R \,n_f$ channel, the failure of the
$z_{\text{cut}}$ condition is now associated with a soft gluon divergence.
In the $z_1\to 0$ soft limit, the spin averaged splitting kernel factorises
into a product of an eikonal factor (associated to the emission of
$g_1$) and a $1\to 2$ splitting function.
In this configuration we obtain:
\begin{multline}\label{eq:failcatr}
	\left(\frac{\theta^2}{\sigma_0}
          \frac{\sd^2\sigma_{\mathcal{R}}^{(2)}}{\sd\theta^2 \, \sd
            z}\right) _{z_1 < z_\text{cut}}^{C_A T_R} = \left(\frac{\alpha_s}{2\pi}\right)^2  C_A T_R n_f  \left(2 \,z (1-z) \right)^{-2\ep}\theta^{-2\epsilon} p_{qg}(z,\ep) \times \\
	\times \left(-\frac{\zc^{-2\ep}}{\ep} \ln \frac{4 }{\theta^2}   - \frac{\pi^2}{6} - \frac12 \ln^2  \frac{4 }{\theta^2}     \right)  \Theta(z - z_{\text{cut}})\,\Theta(1 - z_{\text{cut}}-z) \, .
\end{multline}
This result is similar to what was obtained for quark jets in the
$C_F^2$ channel reported in Eq.~(27) of Ref.~\cite{Anderle:2020mxj},
modulo the replacement $p_{qq} \to p_{qg}$. The final result is
obtained as the sum of the two contributions Eqs.~\eqref{eq:catrfinalreal} and \eqref{eq:failcatr}.

\subsection{The $C_A^2 $ channel}\label{sect:ca2}
As explained in Sec.~\ref{sect:defs}, we work in the region
$\textrm{min}\{\theta_{ij}\} = \theta_{12}$, see
Fig.~\ref{fig:triple-collinear-ca2}. This is completely general as the
other sectors have an identical structure due to the three-fold
symmetry present in the splitting kernel. To account for the full
$1/3!$ symmetry factor, one can simply choose $z_p<1-z_p$. Clearly in
this sector the only collinear singularity arises when
$\theta_{12} \to 0$.
As in the $C_AT_Rn_f$ channel, also in this case we have to consider
two scenarios, i.e. when the gluon at larger angle either passes or
fails the $z_{\text{cut}}$ condition. In the first scenario, the
angular variable to be fixed is $\theta =\theta_{12,3}= \theta_{g}$
now defined as
 \begin{align}\label{eq:thetagdef}
\theta_{g}^2 = \frac{z_1}{z_1 + z_2} \theta^2_{13} + \frac{z_2}{z_1 + z_2} \theta^2_{23} - \frac{z_1 z_2}{(z_1+z_2)^2} \theta^2_{12} \ ,
 \end{align}
 and $z_3 = z$ (with $1-z_{\text{cut}} > z > z_{\text{cut}}$). We obtain:
 \begin{multline}
 \label{eq:fpass}
 	\left(\frac{\theta^2}{\sigma_0} \frac{\sd^2\sigma_{\mathcal{R}}^{(2)}}{\sd\theta^2 \, \sd z}\right)_{z_3 > z_\text{cut}}^{C_A^2} = C_A^2 \left( \frac{\alpha_s}{2\pi} \right)^2   \bigg( \frac{H_{\text{soft-coll.}}(\theta^2,z,\epsilon)}{ \epsilon^2}
 	+ \frac{H_{\mathrm{coll.}}(\theta^2,z,\epsilon)}{\epsilon} \\ +\frac{H_{\mathrm{soft}}(\theta^2,z,\epsilon)}{\epsilon} + H^{C_A^2}_{\text{fin.}}(z) \bigg)  \Theta(z - z_{\text{cut}})\,\Theta(1 - z_{\text{cut}}-z)\, ,
 \end{multline}
where we have symmetrised over $z \leftrightarrow 1-z$. The functions
appearing in the above equation are given by
\begin{align}
  H_{\text{soft-coll.}}(\theta^2,z,\epsilon) &=  \, z^{-2\ep} (1-z)^{-2\ep} \, \left(z^{-2\epsilon} + (1-z)^{-2\epsilon}\right) \theta^{-4\epsilon} 4^{2\epsilon}\, p_{gg}(z)
                                               \left(1 - \frac{\pi^2}{6}\epsilon^2\right) \, , \notag \\
  H_{\text{coll.}}(\theta^2,z,\epsilon) &= - z^{-2\ep} (1-z)^{-2\ep} \left(z^{-2\epsilon} + (1-z)^{-2\epsilon}\right) \theta^{-4\epsilon} 4^{\epsilon}\, p_{gg}(z) \notag \\
                                             &\times \left(-\frac{11}{6}+2\ln 2 +\epsilon \left(-\frac{67}{9}+\frac{2\pi^2}{3}+2 \ln^2 2\right)\right) \, , \\
  H_{\text{soft}}(\theta^2,z,\epsilon) &=  - z^{-2\ep} (1-z)^{-2\ep}
                                         \left(z^{-2\epsilon} +
                                         (1-z)^{-2\epsilon}\right)
                                         \theta^{-4\epsilon}
                                         4^{\epsilon} p_{gg}(z) \left(
                                        2 \ln 2 + h^{\rm poles}_{\rm pass} \, \epsilon \right) \, , \notag
\end{align}
where $h^{\rm poles}_{\rm pass}$ is a numerical constant that reads
\begin{equation}\label{eq:hpolespass}
h^{\rm poles}_{\rm pass}\simeq 1.32644693(2)\,.
\end{equation}
The uncertainty in the above number is on the last quoted digit, which
has be rounded as indicated by the bracket notation. We will use the
same notation for all numerical constants quoted in the following,
unless an error is explicitly quoted.
As expected from the structure of the gluon splitting functions, the
function $H^{C_A^2}_{\text{fin.}}(z)$ has soft divergences as
$z \to 0$ and $z \to 1$ which can be singled out, leading to
\begin{align}\label{eq:Gz}
  H^{C_A^2}_{\text{fin.}}(z) = - \frac{h_{\rm pass}^{\rm fin}}{z(1-z)} + G_{z_3 > z_\text{cut}}(z) \, .
\end{align}
The coefficient of the soft divergences at $z=1$ and $z=0$ in the
above expression is the result of a two-fold integration which can be
performed with very high numerical precision. We obtain
\begin{equation}\label{eq:hfinpass}
h_{\rm pass}^{\rm fin}\simeq 3.31674336(8)\,.
\end{equation}
The quantity $G_{z_3 > z_\text{cut}}(z)$ is fully regular and can be
integrated over $z\in[0,1]$
\begin{align}
	 \int_{0}^{1}dz \, G_{z_3 > z_\text{cut}}(z ) \simeq 16.947\pm
  0.004\, .
\end{align}

\begin{figure}
	\centering
       \includegraphics[width=0.7 \textwidth]{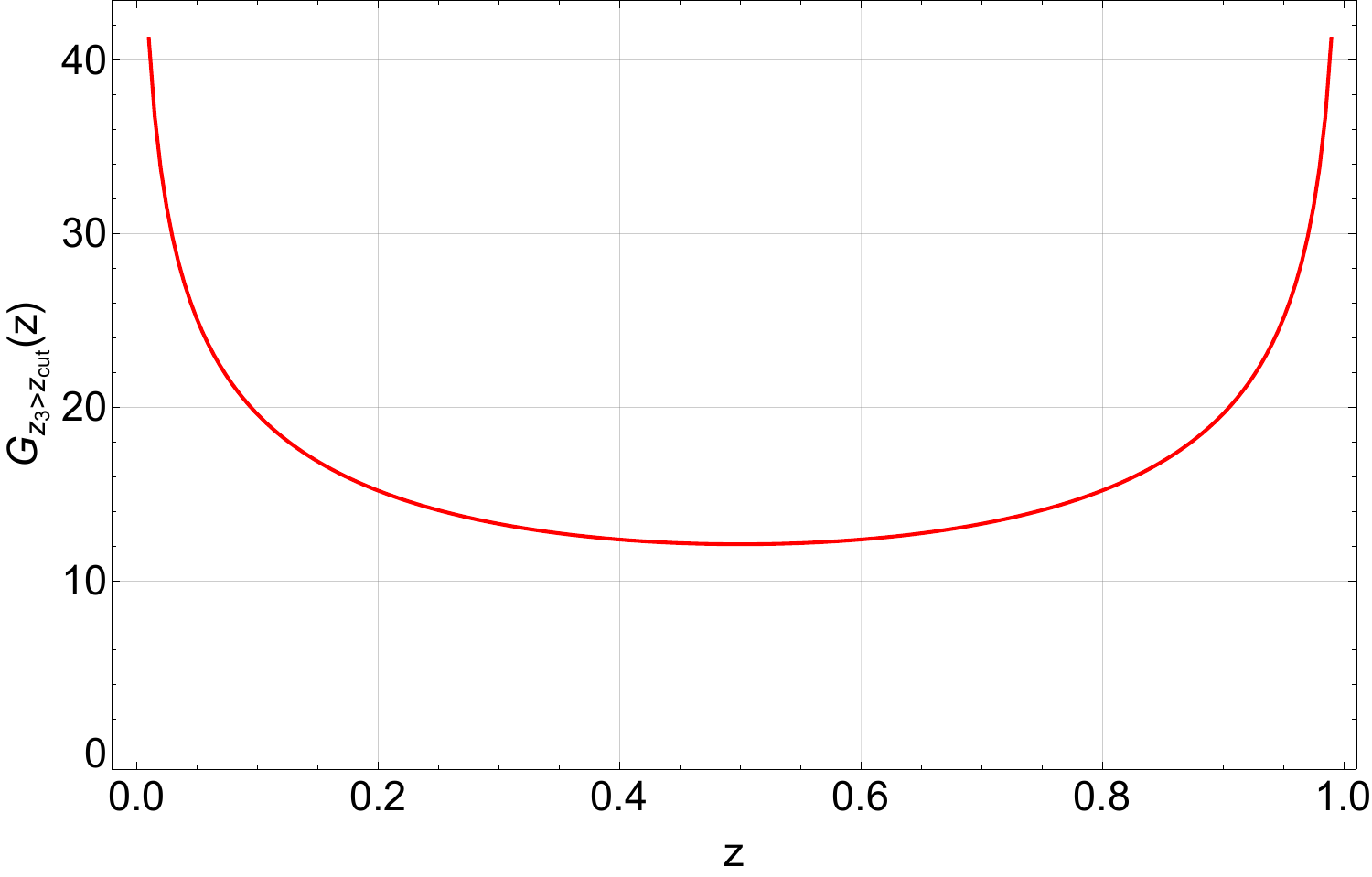}
	\caption{The function $G_{z_3 > z_\text{cut}} (z)$.}
	\label{fig:hfinca2}
\end{figure}

The function $G_{z_3 > z_\text{cut}} (z)$ for this colour channel is
displayed in Fig.~\ref{fig:hfinca2}.
The decomposition in Eq.~\eqref{eq:Gz} will be important later in the
extraction of the NNLL hard collinear coefficient ${\cal B}_2^g(z)$,
which requires a consistent subtraction of the soft contributions.

We now consider the second scenario in which $z_3 <
z_{\text{cut}}$. In the $C_A^2$ channel, this contribution contains a
physical difference from the corresponding result quoted in
Eq.~\eqref{eq:failcatr}. The difference comes from the fact that in
the $z_3\to 0$ limit, the matrix element does not factorise into the
product of independent emissions matrix elements, except for the limit
in which the gluon pair that passes the cut is well separated in angle
from the gluon that fails the cut, i.e. with the notation of
Fig.~\ref{fig:triple-collinear-ca2},
$\theta_{13}\sim \theta_{23}\gg \theta_{12}$.
We can then organise the result as the sum of a term analogous to
Eq.~\eqref{eq:failcatr}, given below
\begin{multline}\label{eq:failca2A}
	\left(\frac{\theta^2}{\sigma_0}
          \frac{\sd^2\sigma_{\mathcal{R}}^{(2)}}{\sd\theta^2 \, \sd
            z}\right) _{z_3 < z_\text{cut}}^{C_A^2,\,(A)} = \left(\frac{\alpha_s}{2\pi}\right)^2  C_A^2  \left(2 \,z (1-z) \right)^{-2\ep}\theta^{-2\epsilon} \,p_{gg}(z) \times \\
	\times \left( -\frac{\zc^{-2\ep}}{\ep} \ln \frac{4 }{\theta^2}   - \frac{\pi^2}{6} - \frac12 \ln^2  \frac{4 }{\theta^2}     \right)  \Theta(z - z_{\text{cut}})\,\Theta(1 - z_{\text{cut}}-z) \, ,
\end{multline}
and a new contribution stemming from the non-independent
(i.e.~correlated) emission contribution. This is evaluated numerically
and obtain
\begin{equation}\label{eq:failca2B}
	\left(\frac{\theta^2}{\sigma_0}
          \frac{\sd^2\sigma_{\mathcal{R}}^{(2)}}{\sd\theta^2 \, \sd
            z}\right) _{z_3 < z_\text{cut}}^{C_A^2,\,(B)} = h^{p_{gg}}_{\rm fail}\,
        p_{gg}(z)\,  \Theta(z - z_{\text{cut}})\,\Theta(1 - z_{\text{cut}}-z) +h^{\delta}_{\rm fail}\,\delta(z)\,,
\end{equation}
where
\begin{equation}\label{eq:fail-const}
  h^{p_{gg}}_{\rm fail}\simeq 0.731081807(5) \, ,\quad
  h^{\delta}_{\rm fail} \simeq  -8.42858916(9)\,.
\end{equation}

\section{Extraction of $\mathcal{B}_2^g(z)$}
\label{sect:b2extraction}

Now we use the results obtained in the previous two sections to
extract the function $\mathcal{B}_2^g(z)$. As we stressed before
$\mathcal{B}_2^g(z)$ reflects the NNLL dynamics that does not arise
from strongly-ordered physics. We follow the procedure outlined in
Sec.~\ref{sect:defs} to extract $\mathcal{B}_2^g(z)$, and after adding
real and virtual corrections for each colour channel we need to remove
pieces pertaining to NLL physics. In analogy with
Eq.~\eqref{eq:quarkP} for quark jets we write down our defining
equation for $\mathcal{B}_2^g(z)$ starting from the NNLL evolution
equation for $G_g(x,t)$
\begin{align}\label{eq:Kvgluon}
G_g(x,t) &= u\, \Delta_g(t) + \int_{t}^{t_0} d t' \int_{z_0}^{1-z_0} d z
  \,\bigg[ 	{\mathcal P}_{gg}(z,\theta')\,G_g(x\,z,t')
  \,G_g(x\,(1-z),t') \notag\\
&+  	{\mathcal P}_{qg}(z,\theta')\,G_q(x\,z,t')
  \,G_q(x\,(1-z),t') \bigg]\frac{\Delta_g(t)}{\Delta_g(t')} + {\mathbb K}_g^{\rm finite}[G_q,G_g] \,,
\end{align}
where ${\mathbb K}_g^{\rm finite}[G_q,G_g]$ is given in
Appendix~\ref{app:kernels} and the inclusive emission probabilities
for gluon jets are given by
\begin{align}\label{eq:gluonP}
	{\mathcal P}_{gg}(z,\theta) &\equiv\,  \frac{C_A}{1-z}\left(1 + \frac{\alpha_s(E^2(1-z)^2\theta^2)}{2\pi}K^{(1)}\right)\notag\\&+\frac{\alpha_s(E^2z^2\theta^2)}{\alpha_s(E^2(1-z)^2\theta^2)}\,\frac{C_A}{z}\left(1 + \frac{\alpha_s(E^2z^2\theta^2)}{2\pi}K^{(1)}\right)\notag \\
	&+{\mathcal B}_1 ^{gg} (z)+\frac{\alpha_s(E^2
   (1-z)^2\theta^2)}{2\pi}\left({\mathcal B}^{gg}_2(z)+{\mathcal B}_1 ^{gg} (z)
   b_0\ln (1-z)^2\right)\,,\notag\\
  {\mathcal P}_{qg}(z,\theta)&\equiv\,{\mathcal B}_1^{qg}(z)+\frac{\alpha_s(E^2
   (1-z)^2\theta^2)}{2\pi}\left({\mathcal B}^{qg}_2(z)+{\mathcal B}_1^{qg}(z)
   b_0\ln (1-z)^2\right)\,,
\end{align}
where the LO anomalous dimensions read
\begin{align}
	{\mathcal B}_1^{gg}(z) = C_A \left(z(1-z) - 2\right) \,,\quad {\mathcal B}_1^{qg}(z) = T_R n_f \left( z^2 + (1-z)^2\right)\, .
\end{align}
The ratio of strong couplings in the second line of
${\mathcal P}_{gg}(z,\theta)$ has the role of restoring the correct
scale of the coupling corresponding to the soft singularity as
$z\to 0$ as opposed to the one at $z\to 1$ that is encoded in the
evolution time~\eqref{eq:time}. This feature is of course present
exclusively for gluon jets, and it is absent in the quark
case~\eqref{eq:quarkP}.
The quantity $\mathcal{B}_2^g(z)$ is then simply the sum of
$\mathcal{B}_2^{gg}(z)$ and $\mathcal{B}_2^{qg}(z)$.
A subtle aspect of the above decomposition between terms that are
interpreted as corrections to either the $g\to gg$ or the $g\to
q\bar{q}$ channel is that they both receive a contribution from the
$C_A T_R$ colour factor. The separation between such contributions to
$\mathcal{B}_2^g(z)$ is not unique and the ambiguity is immaterial as
one can decide to assign the whole correction to either of the two
flavour channels. Conventionally we include it as a correction to
$g\to gg$. We thus write:
\begin{align}
  {\mathcal B}^{gg}_2(z) &\equiv C_A\, T_R\, n_f\, \mathcal{B}_2^{g,\,C_A T_R}(z)+C_A^2\,\mathcal{B}_2^{g,\,C_A^2}(z)\,,\notag\\
  {\mathcal B}^{qg}_2(z) &\equiv T_R^2 n_f^2\,\mathcal{B}_2^{g,\,T_R^2}(z)+C_F\,T_R\,n_f\,\mathcal{B}_2^{g,\,C_F\,T_R}(z)\,.
\end{align}
The Sudakov form factor for gluon jets, given at NLL in
Eq.~\eqref{eq:sudakov-nll}, at NNLL then reads
\begin{align}\label{eq:sud-gluon}
\ln \Delta_g(t) &= - \int_t^{t_0} d t' \int_{z_0}^{1-z_0}d z\,\left( {\mathcal P}_{gg}(z,\theta')+ {\mathcal P}_{qg}(z,\theta')\right)\,.
\end{align}
In the following we carry out the calculation of $\mathcal{B}_2^g(z)$
in each of the above colour channels.

\subsection{The $T_R^2 \, n_f^2$ channel}
This channel is distinct in that it has no double real
contribution. From
Eqs.~\eqref{eq:onelooprenorm},~\eqref{eq:gqq1loop},~\eqref{eq:oneloopborn}
and after subtracting the NLL contribution that emerges from
Eq.~\eqref{eq:theta-z-dist} we obtain
\begin{align}
	\mathcal{B}^{g,T_R^2}_2(z) = \,p_{qg}(z) \left(\frac13 + \frac43 \ln(z(1-z))\right) \, ,
\end{align}
which integrates to
\begin{align}\label{eq:B2tr2}
	\int_0^1 dz\, \mathcal{B}^{g,T_R^2}_2(z)  = -\frac{46}{27}\,.
\end{align}

\subsection{The $C_F T_R n_f$ channel}
The extraction of $\mathcal{B}_2^g(z)$ in this channel is very simple
since it starts at NNLL and hence there is no need to subtract NLL
contributions from the total result. Thus we add real,
Eq.~\eqref{eq:cftfbasic} and its mirror symmetric obtained by the swap
$z\leftrightarrow 1-z$, and virtual, Eq.~\eqref{eq:gqq1loop},
corrections to find
\begin{align}
	\mathcal{B}_2^{g,C_F T_R}(z) =p_{qg}(z) \left(
  \ln^2\left(\frac{z}{1-z}\right) - \frac{\pi^2}{3} + 5 \right) +
  H^{C_F T_R}_{\textrm{fin.}}(z) \, ,
\end{align}
whose integral reads
\begin{align}
	\int_0^1 dz\, \mathcal{B}^{g,C_F T_R}_2(z)  = 1 \, .
\end{align}

\subsection{The $C_A T_R n_f$ channel}
We combine the real and virtual terms
Eqs.~\eqref{eq:catrfinalreal},~\eqref{eq:failcatr},~\eqref{eq:onelooprenorm},~\eqref{eq:gqq1loop},~\eqref{eq:gggo1loop}
and~\eqref{eq:oneloopborn} and then subtract the NLL contribution that
emerges from Eq.~\eqref{eq:theta-z-dist}, to obtain
\begin{multline}
\mathcal{B}_2^{g,C_A T_R}(z)  = -p_{qg}(z)  \left(\ln^2z +\ln^2(1-z)\right)+\frac{1}{9} (28-41 z+41 z^2) \\+ \ln z\left(\frac{4}{3(1-z)}-\frac{26}{3}z^2+8 z-7 \right)+ \ln (1-z) \left(\frac{4}{3z}-\frac{26}{3} (1-z)^2+8(1-z) -7 \right),
\end{multline}

which integrates to
\begin{equation}\label{eq:B2catr}
\int_0^1 dz\, \mathcal{B}_2^{g,C_A T_R}(z) =  \frac{593}{54}-\frac{4\pi^2}{9}.
\end{equation}

Before we discuss the more involved $C_A^2$ channel, it is instructive
to compare the integral of the contributions to $\mathcal{B}_2^{g}(z)$
computed so far to the expectation given in
Eq.~\eqref{eq:expform-gluon}.
In particular, given that the term
$\mathcal{F}_{\text{clust.}}^{C_A^2}$ is a pure $C_A^2$ contribution,
the combination of Eqs.~\eqref{eq:B2tr2},~\eqref{eq:B2catr} allows us
to extract the constant $X^g_{\theta^2}$ corresponding to the
observable $\frac{\theta^2}{\sigma_0}\frac{d^2\sigma}{d\theta^2d z} $
used in the calculation. This yields
\begin{equation}
\label{eq:xtheta-gluon}
X^g_{\theta^2} = - C_A \left(\frac{67}{9}-\frac{2\pi^2}{3}\right)+\frac{23}{9} T_R n_f\,,
\end{equation}
which can be used as a cross check in the $C_A^2$ channel below.

\subsection{The $C_A^2$ channel}
\label{sec:B2ca2}
The procedure for extracting $\mathcal{B}_2(z)$ is the same as in the previous
channel.  The relevant real emission terms are given in
Eqs.~\eqref{eq:fpass},~\eqref{eq:failca2A},~\eqref{eq:failca2B} which
can then be combined with the virtual contributions emerging from
Eqs.~\eqref{eq:onelooprenorm},~\eqref{eq:gggo1loop},~\eqref{eq:oneloopborn}. After
removal of NLL terms the result can be expressed as
\begin{equation}\label{eq:B2ofzca2}
\mathcal{B}_2^{g,C_A^2}(z)  = \mathcal{B}_2^{g,C_A^2,
  \text{analytic}}(z) + G_{z_3>z_{\rm cut}}(z) +  \mathcal{B}_2^{\text{endpoint}}(z).
\end{equation}
The term $\mathcal{B}_2^{g,C_A^2, \text{analytic}}(z) $ is computed
analytically while the contribution $G_{z_3>z_{\rm cut}}(z)$ is
determined numerically. Finally there is an endpoint contribution
which originates purely from the double-soft limit. This includes the
clustering correction as well as the numerically computed
contributions from the region where $z_3 < \zc$. The term
$\mathcal{B}_2^{g,C_A^2, \text{analytic}}(z)$ reads:
\begin{multline}
 \mathcal{B}_2^{g,C_A^2, \text{analytic}}(z)  = p_{gg}(z) \left(-2 \ln z \ln(1-z)-\frac{11}{3} \ln z- \frac{11}{3} \ln(1-z) \right)\\+\frac{11}{3} \left(\frac{\ln z}{z}+\frac{\ln(1-z)}{1-z} \right) 
 -\frac{265}{18} +\frac{\pi^2}{3} (2-z(1-z))+4 \,h_{\rm pass}^{\rm poles}-\frac{44}{3} \ln2+8 \ln^2 2\\-2z(1-z) \left(-
   \frac{67}{18}+h_{\rm pass}^{\rm poles}+2 \ln^2 2-\frac{11}{3} \ln
   2\right)+h_{\rm fail}^{p_{gg}}\,(-2+z(1-z))\,.
\end{multline}
Its integral is straightforward and gives:
\begin{equation}
\int dz \,  \mathcal{B}_2^{g,C_A^2, \text{analytic}}(z) = -\frac{535}{27}+\frac{11 \pi^2}{9}+\frac{11}{3} h_{\rm pass}^{\rm poles}-\frac{121}{9} \ln 2+\frac{22}{3} \ln^2 2-4 \zeta(3)-\frac{11}{6}\,h_{\rm fail}^{p_{gg}}.
\end{equation}
The end-point contribution is simply given by
\begin{equation}\label{eq:B2ca2endpoint}
\mathcal{B}_2^{\text{endpoint}}(z) = h_{\rm fail}^{\delta} \,\delta(z)\,.
\end{equation}
The constants $h_{\rm pass}^{\rm poles}$, $h_{\rm fail}^{p_{gg}}$, and
$h_{\rm fail}^{\delta}$ are defined in
Eqs.~\eqref{eq:hpolespass},~\eqref{eq:fail-const} and the function
$G_{z_3>z_{\rm cut}}(z)$ is defined in Eq.~\eqref{eq:Gz}.
The total integral of $\mathcal{B}_2^{g,C_A^2}(z) $ then yields
\begin{equation}
\int_0^1 dz \, \mathcal{B}_2^{g,C_A^2}(z) \simeq -6.314 \pm 0.004\,.
\end{equation}

As a cross check of the above result we can compare to the expected
integral in Eq.~\eqref{eq:expform-gluon}, where $X_{\theta^2}^g$ is
given in Eq.~\eqref{eq:xtheta-gluon}. For this check we still have to
determine the function $\mathcal{F}_{\text{clust.}}^{C_A^2}$.
For the $C_A^2$ channel our procedure for the extraction of
$\mathcal{B}_2^{g,C_A^2}(z) $ agrees exactly with the mMDT/Soft drop
grooming with the Cambridge/Aachen algorithm, for which
$\mathcal{F}_{\text{clust.}}^{C_A^2}$ is known to stem from the
double-soft limit (within the triple-collinear
approximation). Therefore, due to Casimir scaling it is simple to
relate the clustering correction $\mathcal{F}_{\text{clust.}}^{C_A^2}$
to the known result for quark jets (see e.g.~\cite{Anderle:2020mxj})
(cf. also the discussion in Appendix~\ref{app:clustering}). Replacing
$C_F \to C_A$ in the quark jet result we get the following result the
$C_A^2$ component reads:
\begin{equation}
\label{eq:expclust}
\mathcal{F}_{\text{clust.}}^{C_A^2}= \frac{1}{2}
C^2_A\left(\frac{4\pi}{3} \text{Cl}_2\left(\frac{\pi}{3}\right) + \,h^{ C_A}_{\text{clust.}} \right)\,,
\end{equation}
where
\begin{equation}\label{eq:hclust}
h^{ C_A}_{\text{clust.}} \simeq - 1.16363257(4)\,,
\end{equation}
and $\text{Cl}_2$ denotes the Clausen function.
The overall factor of $1/2$ is multiplied to obtain the result for a
single leg. This gives
\begin{equation}
\label{eq:expform2}
(-\gamma_g^{(2)}+ b_0 X_{\theta^2}^g+ \mathcal{F}_{\text{clust.}})_{C_A^2} \simeq -6.31426325(8)\,C_A^2,
\end{equation}
in good agreement with our result.

We conclude this section with a remark on the endpoint contribution
$\mathcal{B}_2^{\text{endpoint}}(z)$ given in
Eq.~\eqref{eq:B2ca2endpoint}.
As discussed above, this originates from double-soft configurations
and is a consequence of the scheme used here to define $z$ and
$\theta$.
We observe that such endpoint terms mark an important difference to
the case of quark jets (cf. Appendix~\ref{app:B2q}), where separate
definitions of $z$ and $\theta$ can be adopted for correlated and
independent emission contributions. In the quark case these are
separated by colour factors, while in the gluon case they are mixed
together in the $C_A^2$ channel. The same double-soft origin is shared
by the clustering correction~\eqref{eq:expclust}. Terms of double-soft
origin are discussed further in Appendix~\ref{app:clustering}.

\section{Moments of EEC and angularities in groomed jets at NNLL}
\label{sec:applications}
In this section we use the calculations presented in this article to
derive NNLL results for the moments of energy-energy correlation (EEC)
and angularities measured on jets groomed according to the mMDT/Soft
drop ($\beta=0$) procedure~\cite{Dasgupta:2013ihk,Larkoski:2014wba}.
These classes of jet substructure observables have received widespread
attention in the literature, with several applications both at hadron
and lepton
colliders~\cite{Frye:2016aiz,Frye:2016okc,Larkoski:2017bvj,Marzani:2017kqd,CMS:2018vzn,ATLAS:2019mgf,Marzani:2019evv,Kardos:2020gty,Kardos:2020ppl,CMS:2021vsp,Cal:2021fla,Caletti:2021oor,Reichelt:2021svh,Hannesdottir:2022rsl}. As a concrete example we consider the processes $Z\to q\bar{q}$ and
$H\to gg$ to analyse both quark and gluon jets. Specifically, we calculate, for the first time, the groomed fractional moments of EEC, which
are defined as~\cite{Banfi:2004yd}
\begin{align}\label{eq:fcx}
	FC^{\mathcal{H}}_x = \frac{2^{-x}}{E^2} \sum_{i\neq j} E_i E_j |\sin\theta_{ij}|^x  (1-|\cos\theta_{ij}|)^{1-x} \ ,
\end{align}
where the sum runs over all particles within a given hemisphere,
$\mathcal{H} \equiv \mathcal{H}_R $ or
$\mathcal{H} \equiv \mathcal{H}_L $, and $E$ denotes the total energy
in the hemisphere. We also consider the
angularities~\cite{Berger:2003iw} (the corresponding NNLL resummation
in the ungroomed case is given in
Refs.~\cite{Larkoski:2014uqa,Bell:2018gce,Banfi:2018mcq,Procura:2018zpn,Bauer:2020npd})
defined w.r.t. the Winner-Take-All (WTA) axis as
\begin{align}\label{eq:lambdax}
  \lambda^{\mathcal{H}}_x = \frac{2^{1-x}}{E} \sum_i E_i \, |\sin\theta_i|^x (1-|\cos\theta_{i} |)^{1-x} \, ,
\end{align}
where once again the sum involves all particles in a given hemisphere. For both observables the parameter $x$ is constrained by IR
safety to be $x < 2$. 
Finally, we define the observables as follows:
\begin{align}
	FC_x \equiv {\max}\left\{FC^{\mathcal{H}_R}_x,FC^{\mathcal{H}_L}_x\right\}, \, 
	\lambda_x \equiv {\max}\left\{\lambda^{\mathcal{H}_R}_x,\lambda^{\mathcal{H}_L}_x\right\} \, .
\end{align}
A NNLL calculation for WTA angularities for quark jets has been
recently presented in Ref.~\cite{Dasgupta:2022fim},\footnote{Notice
  that here we adopt a different normalisation of the observable,
  compared to Ref.~\cite{Dasgupta:2022fim}, in order to match the
  corresponding hadron collider jet definitions.} and below we present
new results for fractional moments of EEC for quark and gluon jets as
well as for angularities measured on gluon jets.
These results allow for a complete phenomenological analysis of
moments of EEC and angularities on groomed jets at hadron and lepton
colliders.

The master formula for the NNLL cumulative cross section $\Sigma(v)$
for a groomed observable $v$ (that here denotes either a moment of EEC
or an angularity) in the two processes considered here can be easily
derived by applying the GFs method to Eq.~\eqref{eq:dsigma}. In the
following we work in the limit $v\ll z_{\rm cut}\ll 1$, that is
commonly considered for these types of groomed observables.
This regime has the advantage that one can neglect power corrections
in $z_{\rm cut}$ and hence the evolution equations for the GFs can be
solved analytically. Specifically, in this limit we can restrict
ourselves to taking the \textit{soft limit} of the
$ {\mathbb K}_q^{\rm finite}[G_q,G_g]$ and
${\mathbb K}_g^{\rm finite}[G_q,G_g]$ functions (cf.
Eqs.~\eqref{eq:Kvquark},~\eqref{eq:Kvgluon} and
Appendix~\ref{app:kernels}). With a little abuse of notation we have
used the same parameter $z_{\rm cut}$ for the definition of groomed
observables and in the calculation of ${\cal B}^g_2(z)$. However, we
stress that (cf. Sec.~\ref{sect:defs}) ${\cal B}^g_2(z)$ is defined
strictly by taking the limit $z_{\rm cut}\to 0$ while in the
observables considered here one can opt to retain finite $z_{\rm cut}$
effects to improve the accuracy of the calculation. The quantity
${\cal B}^g_2(z)$ we have derived applies to both cases with and
without finite $z_{\rm cut}$ effects, since these would be captured by
the full (numerical) solution of the GFs equations (or equivalently by
a NNLL accurate parton shower algorithm).

Using the evolution equation in
Eqs.~\eqref{eq:Kvquark},~\eqref{eq:Kvgluon}, and following similar
steps to those outlined in
Appendix~\ref{app:SDquark},\footnote{As done in
    Appendix~\ref{app:SDquark}, we can work with the approximation
    $E_p\equiv E$ in the evolution time~\eqref{eq:time} since we
work under the assumption that $z_{\rm cut}\ll 1$.} we obtain the
following NNLL results for quark and gluon jets, respectively
\begin{align}\label{eq:master}
\Sigma^q(v) &= \sigma^{Z\to q\bar{q}} _0\left(1+\frac{\alpha_s(E^2)}{2\pi} C_v^{q(1)}(z_{\rm cut})\right)
  e^{-2 R^q_v(v,z_{\rm cut})}\left(1+\frac{\alpha^2_s(E^2)}{(2\pi)^2}\,2 {\cal F}^q_{\text{clust}}(v)\right)\,,\notag\\
\Sigma^g(v) &= \sigma^{H\to gg} _0\left(1+\frac{\alpha_s(E^2)}{2\pi} C_v^{g(1)}(z_{\rm cut})\right)
  e^{-2 R^g_v(v,z_{\rm cut})}\left(1+\frac{\alpha^2_s(E^2)}{(2\pi)^2}\,2 {\cal F}^g_{\text{clust}}(v)\right)\,,
\end{align}
with $\sigma^{Z\to q\bar{q}, H\to gg} _0$ denoting the leading-order
cross sections.

The quantity $R_v(v,z_{\rm cut})$ denotes the Sudakov radiator, which
can be obtained by integrating the inclusive emission probability in
Eqs.~\eqref{eq:quarkP},~\eqref{eq:gluonP} with the measurement
function of the observables~\eqref{eq:fcx},~\eqref{eq:lambdax} for a
single collinear $a\to b c$ splitting, namely
\begin{equation}\label{eq:measurement-function-v}
\Theta(V_x(z,\theta) - v)\,,
\end{equation}
with $V_x = \{FC_x , \lambda_x \}$ and
\begin{align}\label{eq:obscoll}
FC_x(z,\theta) \simeq &\,  z (1-z) \theta^{2-x} \ \ ,\\
\lambda_x(z,\theta) \simeq &\, \,\text{min}(z,1-z) \theta^{2-x} \, ,
\end{align}
where $z$ is the energy fraction of $b$.
This gives (see also Ref.~\cite{Dasgupta:2022fim}) (with
$f\in\{q,g\}$)
\begin{align}\label{eq:radiator-obs}
  R^f_v(v,z_{\rm cut}) \equiv - g^f_{1}(\lambda_v,\lambda_{\zc}) - g_{2}^f(\lambda_v,\lambda_{\zc}) - h^f_2 (\lambda_v) -\frac{\alpha_s}{\pi} h^f_3(\lambda_v ;B_{2,v}^f)\,,
\end{align}
where $\lambda_v \equiv \alpha_s\beta_0\ln 1/v$,
$\lambda_{z_{\rm cut}} \equiv \alpha_s\beta_0\ln 1/z_{\rm cut}$,
$\alpha_s\equiv \alpha_s(E^2)$, and
\begin{equation}
\beta_0 = \frac{b_0}{2\pi}\,.
\end{equation}
The functions in Eq.~\eqref{eq:radiator-obs} read (we use the notation
$C_R^{q}\equiv C_F$ and $C_R^{g}\equiv C_A$) 
\begin{align}\label{eq:g1} 
g^f_{1}(\lambda_v,\lambda_{\zc})= \frac{C_R^{f}}{\pi \alpha_s \beta_0^2} \lambda_{\zc} \ln\left(1-\frac{2 \lambda_v}{2-x}\right) \,,
\end{align}
\begin{align}
g^f_{2}(\lambda_v,\lambda_{\zc})&= \frac{C_R^f}{\pi \alpha_s \beta_0^2} \lambda_{\zc}^2 \left(\frac{1-2 \lambda_v}{2-x - 2 \lambda_v}\right)  + \frac{C_R^f \beta_1}{\pi \beta_0^3} \lambda_{\zc} \frac{2 \lambda_v + (2-x) \ln \left(1-\frac{2 \lambda_v}{2-x}\right)}{2-x - 2\lambda_v} \notag\\&-\frac{C_R^f}{\pi^2 \beta_0^2} \lambda_{\zc} \frac{\lambda_v\, K^{(1)}} {2-x - 2 \lambda_v}  \,, \label{eq:g2} 
\end{align}
\begin{align}
 h^f_2 (\lambda_v) &=  \frac{ B_{1}^f }{2\pi \beta_0} \ln\left(1-\frac{2 \lambda_v}{2-x}\right) \label{eq:h2}\ \ , \\
h^f_3(\lambda_v ;B_{2,v}^f) &= \frac{ B_{1}^f \,\beta_1}{2 \beta_0^2 (2-x - 2\lambda_v)}\left( (2-x) \ln\left(1-\frac{2 \lambda_v}{2-x}\right) + 2\lambda_v \right) \notag\\
&- \frac{B_{2,v}^f }{2\pi \beta_0 (2-x - 2\lambda_v)} \lambda_v   \label{eq:h3}\,,
\end{align}
where
\begin{align}
	\beta_1 = \frac{17 \,C_A^2 - 10 \,C_A T_R n_f - 6 \,C_F T_R n_f}{24 \pi^2} \,.
\end{align}
For the observables of interest $B_{2,v}^f$ emerges from the
integration of Eqs.~\eqref{eq:quarkP},~\eqref{eq:gluonP} with the
phase space constraint in Eq.~\eqref{eq:measurement-function-v} and it
is given by
\begin{align}\label{eq:B2fcx}
B_{2, FC_x}^q   &= B^q_{2,\theta^2} + C_F \,b_0 \frac{2}{2-x}
                  \left(3-\frac{\pi^2}{3}\right) = -\gamma^{(2)}_q+b_0 X_{FC_x}^q \,,\\
  B_{2, FC_x}^g &= B_{2,\theta^2}^g + \frac{ b_0}{2-x} \left(C_A
                  \left(\frac{67}{9} - \frac{2 \pi^2}{3}\right)-
                  \frac{26}{9} T_R n_f\right) \notag\\
  &= -\gamma^{(2)}_g+b_0 X_{FC_x}^g + \frac{1}{2}
    C_A^2\left(\frac{4\pi}{3} \text{Cl}_2\left(\frac{\pi}{3}\right)+\,h^{C_A}_{\text{clust.}}\right)\, ,
\end{align}
for the fractional moments of EEC, and
\begin{align}\label{eq:B2quarkang}
B_{2,\lambda_x}^q &=B^q_{2,\theta^2} + C_F \,b_0 \frac{(9 - \pi^2 +
                    9\ln 2)}{3 (2 - x)} = -\gamma^{(2)}_q+b_0  X_{\lambda_x}^q\,,\\
B_{2,\lambda_x}^g &=B_{2,\theta^2}^g + \frac{b_0}{2-x} \left( C_A
                    \left(\frac{137}{36} - \frac{\pi^2}{3} + \frac{44
                    \ln2}{12}\right)- T_R n_f \left(\frac{29}{18} +
                    \frac{4 \ln 2}{3}\right) \right) \notag\\
  &= -\gamma^{(2)}_g+b_0  X_{\lambda_x}^g + \frac{1}{2}
    C_A^2\left(\frac{4}{3}\pi \text{Cl}_2\left(\frac{\pi}{3}\right)+\,h^{C_A}_{\text{clust.}}\right)\,,
\end{align}
for the WTA angularities.
The quantities $B_{2,\theta^2}^q$ and $B_{2,\theta^2}^g$ are defined
in Eqs.~\eqref{eq:expform-quark} and~\eqref{eq:expform-gluon} while
$X^f_v$ denotes the observable dependent constants for which we obtain
\begin{align}\label{eq:xfcx}
  X_{FC_x}^q &= C_F \,\frac{2}{2-x} \left(3-\frac{\pi^2}{3}\right) + X^q_{\theta^2}\,,\notag\\
  X_{FC_x}^g &=\frac{1}{2-x} \left(C_A \left(\frac{67}{9} - \frac{2
               \pi^2}{3}\right)- \frac{26}{9} T_R n_f\right)+
               X^g_{\theta^2} \,,
\end{align}
and
\begin{align}\label{eq:xlambdax}
  X_{\lambda_x}^q &= C_F \, \frac{(9 - \pi^2 + 9\ln 2)}{3 (2 - x)}+ X^q_{\theta^2}\,,\notag\\
  X_{\lambda_x}^g &=\frac{1}{2-x} \left( C_A \left(\frac{137}{36} - \frac{\pi^2}{3} + \frac{44 \ln2}{12}\right)- T_R n_f \left(\frac{29}{18} + \frac{4 \ln 2}{3}\right) \right)+ X^g_{\theta^2} \,,       
\end{align}
where the constants $X^q_{\theta^2}$ and $X^g_{\theta^2}$ are given in
Eqs.~\eqref{eq:xtheta-quark},~\eqref{eq:xtheta-gluon}. Finally,
$h^{ C_A}_{\text{clust.}}$ is given in Eq.~\eqref{eq:hclust}.
We note that the result for $B_{2,\lambda_x}^q $ has been previously
obtained in Ref.~\cite{Dasgupta:2022fim}.

The clustering corrections ${\cal F}_{\text{clust.}}(v)$ in
Eq.~\eqref{eq:master} arise from the finite terms
$ {\mathbb K}_q^{\rm finite}[G_q,G_g]$ and
${\mathbb K}_g^{\rm finite}[G_q,G_g]$ in the GF
Eqs.~\eqref{eq:Kvquark},~\eqref{eq:Kvgluon}
(cf. Appendix~\ref{app:kernels}), which give rise to a correction to
the Sudakov in Eq.~\eqref{eq:master} and are given by (see also
Ref.~\cite{Dasgupta:2022fim})
\begin{align}
  {\cal F}^q_{\text{clust.}}(v) &= C_F\left(C_F
                                 \frac{4\pi}{3}
                                 \text{Cl}_2\left(\frac{\pi}{3}\right)
                                 + C_A \,h^{C_A}_{\text{clust.}}
                                 + T_R n_f \,h^{ T_R n_f}_{\text{clust.}} \right) \frac{\ln v}{2-x-2\lambda_v}\,,\notag\\
  {\cal F}^g_{\text{clust.}}(v) &= C_A T_R n_f \,h^{ T_R n_f}_{\text{clust.}} \frac{\ln v}{2-x-2\lambda_v} \,,
\end{align}
where
\begin{equation}\label{eq:hclustnf}
h^{ T_R n_f}_{\text{clust.}} \simeq -1.75559363(5) \,.
\end{equation}
A comment on the difference between quark and gluon jets is in
order. This difference can be traced back to the
projection~\eqref{eq:mapB2} used in the definition of the inclusive
emission probability and hence of ${\cal B}_2^f(z)$. As we discussed
extensively in the paper, we adopt a different projection for quark
and gluon jets due to the more involved structure of collinear
singularities in the latter case. This translates into the final
anomalous dimensions $B_{2,v}^f$ given in
Eqs.~\eqref{eq:B2fcx},~\eqref{eq:B2quarkang}. Here we can see that the
gluon jet result $B_{2,v}^g$, unlike $B_{2,v}^q$, includes a part of
the clustering correction, specifically in the $C_A^2$ channel (see
discussion in Sec.~\ref{sec:B2ca2}).

Finally, we comment on the coefficient functions
$C_v^{f(1)}(z_{\rm cut})$ ($f=\{q,g\}$) in
Eq.~\eqref{eq:master}. These are matching coefficients between the
collinear approximation of the generating functionals and the full
${\cal O}(\alpha_s)$ calculation in the limit $v\ll z_{\rm cut}\ll
1$. For the processes and observables considered here they have the
general structure~\cite{Banfi:2018mcq} 
\begin{align}\label{eq:c1obs}
C_v^{q(1)}(z_{\rm cut}) &= H^{q(1)} - 2 X_v^q + C_F \left(8 \ln 2\ln \zc+ \, 6 \ln 2 -\frac{\pi^2}{3}\right)\,,\notag\\
C_v^{g(1)}(z_{\rm cut}) &=H^{g(1)} -  2 X_v^g + C_A \left(8 \ln 2\ln \zc -\frac{\pi^2}{3}\right)\,,
\end{align}
where $X_{v}^f$ are the observable dependent constants of
hard-collinear origin given in Eqs.~\eqref{eq:xfcx},
\eqref{eq:xlambdax}, and the remaining constants are the (only)
process-dependent ingredients of the calculation. Specifically,
$H^{f(1)}$ accounts for the hard-virtual corrections at one loop order
\begin{align}
  H^{q(1)} &= C_F\,\left(\pi^2 - 8 \right)\, , \notag\\
  H^{g(1)} &=   -3\,C_F+C_A\,\left(\pi^2+5\right)   \,.
\end{align}
The presence of $X_v^f$ both in Eqs.~\eqref{eq:c1obs} and in
Eqs.~\eqref{eq:B2fcx},~\eqref{eq:B2quarkang} indicates that the GF
solution is defined in a resummation scheme in which the running of
$\alpha_s$ multiplying the constants of collinear origin is absorbed
into the anomalous dimensions which define the inclusive emission
probability. In alternative approaches to this class of resummations
(e.g. that of
Refs.~\cite{Banfi:2014sua,Banfi:2018mcq,Dasgupta:2022fim}) the
coupling multiplying these constants is evaluated at the
observable-dependent collinear (low) scale, which would remove $X_v^f$
from Eqs.~\eqref{eq:B2fcx},~\eqref{eq:B2quarkang}. The physical result
at a given logarithmic order is of course resummation-scheme invariant
(cf. also footnote~\ref{foot:scheme} in Appendix~\ref{app:SDquark}).

For quark jets, as a check of our results, we compare the resummed predictions for
$FC_x$ and $\lambda_x$ against \texttt{Event2} in Appendix~\ref{app:event2comparison}, and against the formalism of
Refs.~\cite{Banfi:2014sua,Banfi:2018mcq} adapted to groomed
observables in Refs.~\cite{Anderle:2020mxj,Dasgupta:2022fim}.
Specifically, as observed in Ref.~\cite{Dasgupta:2022fim}, in the case
of groomed quark jets the integrated quantities $B_{2,\lambda_x}^q $
and $B_{2, FC_x}^q $ can be extracted from Sec.~3 of
Ref.~\cite{Banfi:2018mcq}, by combining the endpoint of the two loop
DGLAP anomalous dimension $\gamma^{(2)}_f$ with the running of the
one-loop hard-collinear constant $C^{(1)}_{\rm hc}$ and the recoil
correction $\delta{\cal F}_{\rm rec}$,\footnote{The hard-collinear
  correction $\delta{\cal F}_{\rm hc}$ would also contribute for
  ungroomed angularities and fractional moments of EEC.}  which agrees
with our findings in Eqs.~\eqref{eq:B2fcx},~\eqref{eq:B2quarkang}.
Similarly, an independent calculation of the clustering corrections
for quark angularities can be found in Ref.~\cite{Dasgupta:2022fim}.
We also carried out analogous checks for the gluonic results, based on
a straightforward extension of the above formalism to $H\to gg$.
In particular, we checked that the result for $FC_0$, which coincides with the heavy hemisphere mass, agrees with the  result for the latter of Ref.~\cite{Frye:2016aiz}.
We reiterate that, due to the process-independence of the collinear
limit, our results can be used for hadron-collider jets after
supplying the appropriate process-dependent analogues of the constants
in Eqs.~\eqref{eq:c1obs}.

\section{Conclusions and Outlook}
\label{sec:conclusions}
In this paper we have presented a generating functional formulation of
the NNLL resummation of collinear logarithms produced by multiple
timelike collinear parton splittings.
In particular, we have addressed one of the key elements of the
generating functionals method, namely the Sudakov form factor, which
appears in the formalism as the no-branching probability for a given
fragmenting parton.
We have pointed out that for general NNLL accuracy in the collinear
limit, alongside correcting the real emission matrix-element using the
triple-collinear splitting functions, the Sudakov form factor also
needs to be augmented to the two-loop order. Here we focussed on gluon
fragmentation, and have provided the necessary extension by computing
the anomalous dimension ${\cal B}^g_2(z)$ which governs the intensity
of collinear radiation off a gluon as a function of a suitably defined
longitudinal momentum fraction $z$.
We envisage that the results presented here will be instrumental to
address the following classes of resummation problems:
\begin{enumerate}
\item Problems that do not admit a closed-form solution such as, for
  instance, the differential fragmentation of a jet into a system of
  microjets. This class of problems is sensitive to the whole
  non-linear structure of the generating functionals evolution
  equations provided in this article, which reflects the recursive
  nature of collinear fragmentation. As such, in the general case only
  a numerical solution of these equations can be envisioned, which can
  be achieved either using discretisation techniques or Monte Carlo
  methods. In particular, our results are necessary for the extension
  to NNLL of the NLL results of Ref.~\cite{Dasgupta:2014yra}. We will
  address some of these aspects in a future
  publication~\cite{vanBeekveld:2023?} together with an algorithmic
  solution to the problem of collinear fragmentation.

\item The second class of collinear problems that can be addressed
  with the formalism outlined here is that of semi-inclusive
  observables that are not sensitive to the full non-linear structure
  of the evolution equations, which can now be solved with analytic
  methods~\cite{Frye:2016aiz,Frye:2016okc,Larkoski:2017bvj,Marzani:2017kqd,Marzani:2019evv,Kardos:2020gty,Kardos:2020ppl,Caletti:2021oor,Reichelt:2021svh,Hannesdottir:2022rsl}. Examples
  of these problems, considered in this article, are groomed event
  shapes and moments of EEC. We derived new NNLL results for a family
  of groomed angularities defined w.r.t. the Winner-Take-All axis and
  fractional moments of energy-energy correlators, which open up a
  range of interesting phenomenological studies at hadron colliders
  such as the LHC.

\item Another interesting direction is the extension to more general
  observables that also exhibit sensitivity to soft physics, including
  the important cases of ungroomed global event shapes and non-global
  observables (in the latter case a GFs formulation is given in
  Refs.~\cite{Banfi:2021xzn,Banfi:2021owj}). Although many of the
  above NNLL resummations (e.g. for global event shapes) are known
  from the literature, their formulation in the GFs language would
  pave the way for a single resummation tool capable of achieving NNLL
  accuracy across a wide class of observables. Similar considerations
  apply to the space-like collinear branching of initial-state
  partons, which will be a crucial future step.
  Another key consideration will be the inclusion of spin correlations
  at NNLL via the use of polarised splitting kernels in the GF
  evolution equations.
\end{enumerate}

An important additional aspect of the formalism presented here is that
it allows one to formulate collinear resummation in a language that
resembles that of parton showers, hence offering insight on the design
of future NNLL algorithms.
For instance, the Sudakov form factor calculated in this article
constitutes a crucial ingredient to reach this perturbative accuracy
and it is therefore a key element of future NNLL parton showers.
Finally, the results derived in this article are distributed in Mathematica
format with the arXiv submission of this paper.


\section*{Acknowledgements}
We thank Ludovic Scyboz, Gr\'egory Soyez and Gavin Salam for useful
discussions through the course of this work and comments on the
manuscript. Furthermore, we are grateful to Gavin Salam for
collaboration on the related forthcoming
Ref.~\cite{vanBeekveld:2023?}. We would also like to thank other
colleagues on the PanScales collaboration whose thoughts and efforts
towards pushing forward the logarithmic accuracy of parton showers
have directly motivated this work. This work has been partly funded by
the European Research Council (ERC) under the European Union's Horizon
2020 research and innovation program (grant agreement No 788223) (MvB,
MD, BKE and JH) and by the U.K.'s Science and Technologies Facilities
Council under grant ST/T001038 (MD). The work of PM is funded by the
European Union (ERC, grant agreement No. 101044599, JANUS). Views and
opinions expressed are however those of the authors only and do not
necessarily reflect those of the European Union or the European
Research Council Executive Agency. Neither the European Union nor the
granting authority can be held responsible for them.

\appendix

\section{Leading order splitting functions}
\label{app:SF}
In this appendix we report the well known expression of the
unregularised tree-level splitting kernels used in the main text. We
factor out the colour factors using the notation
\begin{align}
P_{qq}(z,\epsilon)=C_F\, p_{qq}(z,\epsilon)\,,\quad P_{gg}(z)=C_A\, p_{gg}(z)\,,\quad P_{qg}(z,\epsilon)=T_R \,n_f\, p_{qg}(z,\epsilon)\,,
\end{align}
where the splitting functions in $4-2\epsilon$ dimensions read
\begin{align}
  p_{qq}(z,\epsilon) &=\frac{1+z^2}{1-z}-\epsilon (1-z)\,,\notag\\
  p_{gg}(z) &=\frac{1}{1-z}+\frac{1}{z}-2+z(1-z)\,,\notag\\
  p_{qg}(z,\epsilon) &=\frac{z^2+(1-z)^2-\epsilon}{1-\epsilon}\,.
\end{align}
For the sake of simplicity, in the text we also use the following
notation for the $\epsilon=0$ case
\begin{align}
  p_{qq}(z)&\equiv p_{qq}(z,\epsilon=0)\,,\notag\\
  p_{qg}(z)&\equiv p_{qg}(z,\epsilon=0)\,.
\end{align}

\section{Expression of ${\mathcal B}_2(z)$ for quark fragmentation}
\label{app:B2q}
The expression of
${\mathcal B}^q_2(z)$ can be organised as follows:\footnote{The
  function ${\mathcal B}^q_2(z)$ computed in
  Ref.~\cite{Dasgupta:2021hbh} is defined as the ${\mathcal B}^q_2(z)$
  used here multiplied by $\left(\alpha_s/(2\pi)\right)^2$.}
\begin{align}
	 {\mathcal B}^q_2(z) &=
                                       C^2_F\,
  \mathcal{B}_2^{q,C_F^2}(z) +C_F C_A\, \mathcal{B}_2^{q,\,C_F C_A}(z) +C_F T_R n_f \,
  \mathcal{B}_2^{q,\,C_F T_R}(z) + C_F\left(C_F - \frac{C_A}{2}\right)
  \mathcal{B}_2^{q,\,\textrm{id.}}(z)  \ .
\end{align}
The above functions read
\begin{align}
	\mathcal{B}_2^{q,\,\textrm{id.}} (z) &= 4z-\frac{7}{2} +\frac{5z^2-2}{2(1-z)} \ln z+\frac{1+z^2}{1-z}\left(\frac{\pi^2}{6} -\ln z \ln (1-z)-\text{Li}_2(z)\right)\,, \\
	\mathcal{B}_2^{q,\,C_F T_R} (z) &= - b_0^{(n_f)} p_{qq}(z) \ln z + b_0^{(n_f)} (1-z) - K^{(1),n_f} (1+z) + 2\,b_0^{(n_f)} (1+z) \ln(1-z)\, , \\ 
	\mathcal{B}_2^{q,\,C_F C_A}(z) &= - b_0^{(C_A)} p_{qq}(z) \ln z + b_0^{(C_A)} (1-z) +\frac{3}{2} \frac{z^2 \ln z}{1-z} +\frac12 (2z-1)\\& \hspace{-1.7cm}+2 \,b_0^{(C_A)} (1+z) \ln(1-z) + p_{qq}(z) \left(\ln^2z+
   \mathrm{Li}_2\left(\frac{z-1}{z}\right)+2\,
   \mathrm{Li}_2(1-z)\right) - K^{(1),C_A} (1+z) \,, \notag\\
\mathcal{B}_2^{q,C_F^2} (z) &= p_{qq}(z) \left(-3 \ln z -2 \ln z \ln(1-z) +2\, \text{Li}_2\left(\frac{z-1}{z}\right)\right)  - 1 + H^{\rm fin.}(z)\,,
\end{align}
where $H^{\rm fin.}(z)$ is given by a 1-fold integral ({\em cf.}
Figure~4 of ref.~\cite{Dasgupta:2021hbh}), that is provided in
Mathematica format as an ancillary file with the arXiv preprint of
this article.
The function $ {\mathcal B}^q_2(z)$ is regular in the soft limit
$z \to 1$ and is thus fully integrable over $z \in [0,1]$.

\section{The NNLL ${\mathbb K}^{\rm finite}$ kernel}
\label{app:kernels}
In this appendix we report the functions ${\mathbb K}_f^{\rm finite}$
(with $f\in \{q,g\}$) entering the NNLL evolution equation for the
quark generating functionals given in
Eqs.~\eqref{eq:Kvquark},~\eqref{eq:Kvgluon}.
We will start by presenting the result for quark jets due to its
simpler structure, and later give the gluon counterpart.

\subsection{Quark fragmentation}
We can express ${\mathbb K}_q^{\rm finite}$ as a difference between
two terms which encode the (subtracted) real corrections to the first
of Eqs.~\eqref{eq:GF-NLL} and its double counting with the iteration
of the NLL kernel, respectively. That is:
\begin{equation}
 {\mathbb K}_q^{\rm finite}[G_q,G_g]\equiv {\mathbb K}_q^{\rm R}[G_q,G_g]-{\mathbb K}_q^{\rm DC}[G_q,G_g]\,.
\end{equation}
The difference of $ {\mathbb K}_q^{\rm R}$ and ${\mathbb K}_q^{\rm
  DC}$ ensures that the quantity ${\mathbb K}_q^{\rm finite}$ is infrared
finite and purely NNLL.

\begin{figure}[h]
	\centering
	\begin{tikzpicture}[scale=2.5] 
		
		\coordinate (bq1) at (0,0);
		\coordinate (bq2) at (1,0); 
		\coordinate (bq3) at (2.5,0);
		
		\coordinate (bg1) at (1.2,0);
		\coordinate (eg1) at (1.5,0.6);
		\coordinate (bg2) at (1.5,0.6);
		\coordinate (eg2) at (1.3,1.0);
		\coordinate (bg3) at (1.5,0.6);
		\coordinate (eg3) at (1.9,0.8);
		
		\draw [quark] (bq1) -- (bq2);
		\draw [gluon] (bg1) -- (eg1) ;
		\node at (1.4,0.1) {\small \,\,\,\,\,$\theta_{12,3}$};
		\draw [quark] (bg2) -- (eg2) node [pos=1,left] {\small $z_1=(1-z)z_p$};
		\node at (1.58,0.8) {\small $\theta_{12}$};
		\draw [quark] (bg3) -- (eg3) node [pos=1,right] {\small $z_2=(1-z)(1-z_p)$};
		\draw [quark] (bq2) -- (bq3) node
		[pos=1,right] {\small $z_3 = z$} ;

	\end{tikzpicture}
	\caption{The diagram representing gluon decay to a $q\bar{q}$ pair, where the quark from the gluon decay is either identical or non-identical to the initiating quark.}\label{fig:a-splits}
	\label{fig:a-splits}
\end{figure}
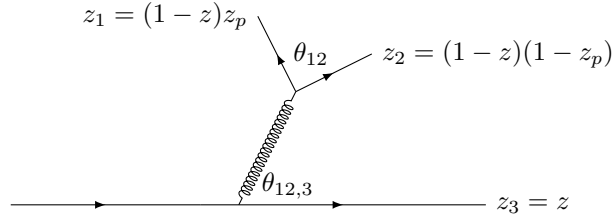

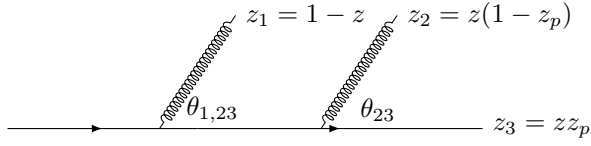
\begin{figure}[h]
	\centering
	\begin{tikzpicture}[scale=2.5] 
		
		\coordinate (bq1) at (0,0);
		\coordinate (bq2) at (1,0); 
		\coordinate (bq3) at (2.5,0);

		\coordinate (bg1) at (.8,0);
		\coordinate (eg1) at (1.2,0.6);
		\coordinate (bg2) at (1.65,0);
		\coordinate (eg2) at (2.05,0.6);
		
		\draw [quark] (bq1) -- (bq2);
		\draw [gluon] (bg1) -- (eg1) node [pos=1,right] {\small $z_1=1-z$} ;
		\node at (1.02,0.1) {\small \,\,\,\,\,$\theta_{1,23}$};
		\draw [gluon] (bg2) -- (eg2) node [pos=1,right] {\small $z_2=z(1-z_p)$};
		\node at (1.95,0.1) {\small $\theta_{23}$};
		\draw [quark] (bq2) -- (bq3) node
		[pos=1,right] {\small $z_3 = z z_p$};

	\end{tikzpicture}
        \caption{The diagram representing the gluon emission $C_F^2$
          channel.}\label{fig:b-splits}
\label{fig:b-splits}
\end{figure}
Following the definition of the inclusive emission
probability~\eqref{eq:quarkP} we obtain:
 \begin{align}\label{eq:KR}
 {\mathbb K}_q^{\rm R}&[G_q,G_g] = \sum_{(A)} \frac{1}{S_2}\int d\Phi^{(A)}_{3}\, P^{(A)}_{1\to 3} \,\bigg\{G_{f_1}(x\,z_p\,(1-z),t_{1,2})
  \,G_{f_2}(x\, (1-z_p)\,(1-z), t_{1,2}) \notag\\ & \times G_q(x\, z, t_{12,3})
   - G_{f_{12}}(x\, (1-z), t_{12,3}) \,G_q(x\, z,t_{12,3}) \bigg\}\frac{\Delta_q(t)}{\Delta_q(t_{1,2})}\Theta(t_{12,3} - t)\notag\\
 & +\int d\Phi^{(B)}_{3}\, P^{(B)}_{1\to 3} \,\bigg\{G_g(x\,(1-z),t_{1,23})
  \,G_g(x\, z \,(1-z_p), t_{2,3})\, G_q(x\, z\,z_p, t_{2,3}) \notag\\
  &- G_{g}(x\, (1-z),
    t_{1,23}) \,G_q(x\, z,t_{1,23})
    \bigg\}\frac{\Delta_q(t)}{\Delta_q(t_{2,3})}\,\Theta(t_{2,3}-t_{1,3})\Theta(t_{1,23} - t)\,,
 \end{align}
 where we have used the notation $t_{i,j}$ to indicate the value of
 the evolution time~\eqref{eq:time} corresponding to the angle
 $\theta_{i,j}$, and $E_p$ is set to the energy of the
 parent parton in the $1\to 3$ branching.
 Moreover, we have parameterised the integrands as~\footnote{We notice
   that the production of a $q \bar{q} q$ final state with identical
   flavours contributing to the $C_F(C_F-C_A/2)$ is finite, and
   therefore does not factorise into the product of two splitting
   functions in the strong angular ordered limit.}
\begin{align}\label{eq:Asplits}
P_{1\to 3}^{(A)} &\equiv \frac{(8\pi)^2}{s_{123}^2}\alpha^2_s(E_p^2g^2(z)\,\theta_{12,3}^2)
                   \langle \hat{P}\rangle _{\tiny{C_F C_A\,,C_F T_R n_f\,,C_F(C_F-C_A/2)}}\,,\\
 P_{1\to 3}^{(B)} &\equiv \frac{(8\pi)^2}{s_{123}^2}\alpha^2_s(E_p^2g^2(z)\,\theta_{1,3}^2)
   \langle \hat{P}\rangle _{\tiny{C_F^2}}\,.
\end{align}
 Here, $P_{1\to 3}^{(A)} $ is parameterised according to the phase
 space depicted in Fig.~\ref{fig:a-splits}, while $P_{1\to 3}^{(B)} $
 is parameterised according to Fig.~\ref{fig:b-splits}. The
 corresponding phase space measures, denoted by $d\Phi^{(A,B)}_{3}$
 in~\eqref{eq:KR} are obtained from Eq.~\eqref{eq:psnormal} by setting
 $\epsilon=0$ and performing the change of variable (with the
 corresponding Jacobian) of
 Figs.~\ref{fig:a-splits},~\ref{fig:b-splits}.
 The sum in Eq.~\eqref{eq:KR} runs over all the $(A)$ colour channels
 of the type $q\to f_1 f_2 q (\bar{q})$ defined in
 Eq.~\eqref{eq:Asplits}, and $f_i$ denotes the flavour of parton
 $i$. The factor $1/S_2$ is the symmetry factor to be applied in the
 case of identical particles (notably $S_2=2!$ in the $C_F C_A$ and in
 the $C_F(C_F-C_A/2)$ channels).

 The phase space integrals are now intended to be regulated by the
 usual cutoff procedure used so far, that implies an upper cut $t_0$
 on evolution time (i.e. a lower cut on angles) and a cut on the
 energy fractions $1-z_0 \geq \{z,z_p\} \geq z_0$. All the
 calculations are then meant to be performed by taking the limit of
 $t_0\to \infty$ and $z_0\to 0$ after combining
 $ {\mathbb K}_q^{\rm R}$ and ${\mathbb K}_q^{\rm DC}$.
 We observe that each term in curly brackets in
 ${\mathbb K}_q^{\rm R}$ is obtained by subtracting from each real
 configuration a differential counterterm defined by projecting the
 real kinematics into a given underlying Born phase space
 point. Crucially, this requires an infrared-and-collinear-safe
 definition of the branching variables $\theta$ and $z$, that uniquely
 specifies the definition of $ {\mathcal B}^q_2(z)$.
 Therefore, the scheme that defines $ {\mathcal B}^q_2(z)$ is tightly
 connected to the specific form of
 ${\mathbb K}_q^{\rm finite}[G_q,G_g]$ in Eq.~\eqref{eq:Kvquark},
 which specifically adopts the definition of $z$ and $\theta$ of
 Ref.~\cite{Dasgupta:2021hbh} outlined in Sec.~\ref{sec:B2quark}. An
 alternative scheme will modify the expression of both
 ${\mathcal B}^q_2(z)$ and ${\mathbb K}_q^{\rm finite}[G_q,G_g]$ by
 modifying the precise definition of $\theta$ and $z$, but the
 prediction of the evolution equation is clearly invariant under any
 scheme change of this type.
 Following similar considerations for the double-counting term we find
  \begin{align}\label{eq:KDC}
 {\mathbb K}_q^{\rm DC}&[G_q,G_g] =
                         \sum_{g\to f\bar{f}}
    \int_t^{t_0} d t_{12,3} \,d t_{1,2}\int_{z_0}^{1-z_0} d z \,d z_p P_{qq}(z)
    P_{f g}(z_p) \,\bigg\{G_{f}(x\,z_p\,(1-z),t_{1,2})\notag\\
&\times G_{f}(x\, (1-z_p)\,(1-z), t_{1,2})  G_q(x\, z, t_{12,3})
   - G_{g}(x\, (1-z), t_{12,3}) \,G_q(x\, z,t_{12,3}) \bigg\}\notag\\&\times\frac{\Delta_q(t)}{\Delta_q(t_{1,2})}\,\Theta(t_{1,2} -t_{12,3})\notag\\
 & +\int_t^{t_0} d t_{1,23} \,d t_{2,3}\int_{z_0}^{1-z_0} d z \,d z_p P_{qq}(z)
    P_{qq}(z_p) \,\bigg\{G_g(x\,(1-z),t_{1,23})
  \,G_g(x\, z \,(1-z_p), t_{2,3}) \notag\\
  &\times G_q(x\, z\,z_p, t_{2,3}) - G_{g}(x\, (1-z),
    t_{1,23}) \,G_q(x\, z,t_{1,23})
    \bigg\}\frac{\Delta_q(t)}{\Delta_q(t_{2,3})}\,\Theta(t_{2,3} -t_{1,23})\,,
 \end{align}
 where the sum runs over the $g\to gg$ and $g\to q\bar{q}$ splitting
 channels.
 Eq.~\eqref{eq:KDC} is obtained by calculating the contribution to
 Eq.~\eqref{eq:KR} due to the iteration of the NLL kernel in
 Eq.~\eqref{eq:GF-NLL}. We notice that the difference between
 Eqs.~\eqref{eq:KR} and~\eqref{eq:KDC} (i.e.
 ${\mathbb K}_q^{\rm finite}[G_q,G_g]$) only contributes in the regime
 where all angles are commensurate
 $t_{2,3} \simeq t_{1,23} \simeq t_{1,3}$. Conversely,
 ${\mathbb K}_q^{\rm finite}[G_q,G_g]$ vanishes in the strongly
 ordered limit. This implies that one can always approximate the
 angles in the Sudakov and in the GFs according to
 $t_{2,3} \simeq t_{1,23} \simeq t_{12,3}\simeq t_{1,3}$ neglecting
 higher-order, N$^3$LL terms. This property proves useful 
 when performing a (semi-)analytic calculation using this formalism. 
 
\subsection{Gluon fragmentation}
We now extend the result of the previous section to the case of gluon
jets. The derivation proceeds through analogous arguments to the quark
case, with two main differences. The first difference is related to
the scheme for the anomalous dimension $ {\mathcal B}^g_2(z)$ used in
the main text for its computation. As we discussed at length in
Sec.~\ref{sect:defs}, the definition of $ {\mathcal B}^g_2(z)$ relies
on a specific projection~\eqref{eq:mapB2} from the three-particle to
the two-particle phase space, which is reflected in definition of the
inclusive emission probability as well as in the definition of $z$ and
$\theta$ used in the calculation of ${\mathbb K}^{\rm
  R}[G_q,G_g]$. The latter now reads:
\begin{equation}\label{eq:KRgluon}
  {\mathbb K}_g^{\rm R}[G_q,G_g] = {\mathbb K}_g^{{\rm R},\, C_A\,T_R}[G_q,G_g] +  {\mathbb K}_g^{{\rm R},\, C_F\,T_R }[G_q,G_g] + {\mathbb K}_g^{{\rm R},\, C_A^2}[G_q,G_g] \,,
\end{equation}
where (using the parameterisation of
Figs.~\ref{fig:triple-collinear-nonab} and~\ref{fig:triple-collinear-ab})
\begin{align}
{\mathbb K}_g^{{\rm R},\, C_A\,T_R}&[G_q,G_g] =\int d\Phi^{(A)}_{3}\, P^{C_A\,T_R}_{1\to 3} \,\bigg\{G_{q}(x\,z_p\,(1-z),t_{2,3})
  \,G_{q}(x\, (1-z_p)\,(1-z), t_{2,3}) \notag\\ & \times G_g(x\, z, t_{1,23})
   - G_{g}(x\, (1-z), t_{1,23}) \,G_g(x\, z,t_{1,23}) \bigg\}\frac{\Delta_g(t)}{\Delta_g(t_{2,3})}\Theta(t_{1,23} - t)\,,
\end{align}
\begin{align}
{\mathbb K}_g^{{\rm R},\, C_F\,T_R}&[G_q,G_g]=\int d\Phi^{(B)}_{3}\, P^{C_F\,T_R}_{1\to 3} \,\Bigg[\bigg\{G_q(x\,(1-z),t_{2,13})
  \,G_g(x\, z \,(1-z_p), t_{1,3}) \notag\\
  &\hspace{-1.2cm} \times G_q(x\, z\,z_p, t_{1,3}) - G_{q}(x\, (1-z),
    t_{2,13}) \,G_q(x\, z,t_{2,13})
    \bigg\}\frac{\Delta_g(t)}{\Delta_g(t_{1,3})}\,\Theta(t_{1,3} -t_{1,2})\Theta(t_{2,13} - t)\notag\\
 & +\{2\leftrightarrow 3,\,z\leftrightarrow 1-z\}\bigg]\,.
\end{align}
Finally, for the $C_A^2$ channel we must distinguish between the case
in which $z > z_{\rm cut}$ and $z < z_{\rm cut}$, following the scheme
used to define ${\cal B}_2^g(z)$. However, since the latter is defined
in the $z_{\rm cut}\to 0$ limit, we can entirely neglect the region
$z < z_{\rm cut}$ in the calculation of
${\mathbb K}_{g}^{{\rm R},\, C_A^2}[G_q,G_g]$.
Using the phase space parameterisation of
Fig.~\ref{fig:triple-collinear-ca2} we then find
\begin{align}\label{eq:permutations-gluons}
{\mathbb K}_{g}^{{\rm R},\, C_A^2}[G_q,G_g] = \frac{1}{3}\left({\mathbb
  K}_{g,\{12,3\}}^{{\rm R},\, C_A^2}[G_q,G_g]+\{{\rm 2\,cyclic\,permutations}\}\right)\,,
\end{align}
where we defined
 \begin{align}
{\mathbb K}_{g,\{12,3\}}^{{\rm R},\, C_A^2}&[G_q,G_g] =\frac{1}{2!}\int d\Phi^{(A)}_{3}\, P^{C_A^2}_{1\to 3} \,\bigg\{G_{g}(x\,z_p\,(1-z),t_{1,2})
  \,G_{g}(x\, (1-z_p)\,(1-z), t_{1,2}) \notag\\ & \times G_g(x\, z, t_{12,3})
   - G_{g}(x\, (1-z), t_{12,3}) \,G_g(x\, z,t_{12,3}) \bigg\}\notag \\
   &\times \frac{\Delta_g(t)}{\Delta_g(t_{1,2})}\, \Theta(t_{1,2}-\max\{t_{1,3}, t_{2,3}\})\Theta(t_{12,3} - t) \,.
 \end{align}
 The additional combinatorial factor $1/2!$ accounts for the remaining
 two-fold symmetry in the $g\to ggg$ splitting function, while the
 factor $1/3$ in Eq.~\eqref{eq:permutations-gluons} can be removed by
 choosing one of the permutations.

The second important difference between the gluon and quark cases
concerns the double counting term ${\mathbb K}^{\rm DC} [G_q,G_g]$,
which in the gluonic case is more involved due to the fact that two
different splitting channels contribute to the NLL evolution equation
for $G_g$ (cf. Eq.~\eqref{eq:GF-NLL}).
We write it as
\begin{equation}\label{eq:KDCgluon}
  {\mathbb K}_g^{\rm DC}[G_q,G_g] = {\mathbb K}_g^{{\rm DC},\, C_A\,T_R}[G_q,G_g] +  {\mathbb K}_g^{{\rm DC},\, C_F\,T_R }[G_q,G_g] + {\mathbb K}_g^{{\rm DC},\, C_A^2}[G_q,G_g] \,,
\end{equation}
where (using the parameterisation of
Figs.~\ref{fig:triple-collinear-nonab},~\ref{fig:triple-collinear-ab}
and~\ref{fig:triple-collinear-ca2}, respectively)\footnote{We note
  that the $C_A\,T_R$ contribution to the double counting term
  $ {\mathbb K}_g^{{\rm DC}, \, C_A\,T_R}[G_q,G_g] $ is symmetric in
  $z\leftrightarrow 1-z$, while
  ${\mathbb K}_g^{{\rm R},\, C_A\,T_R}[G_q,G_g]$ given above is
  not. This symmetrisation in
  ${\mathbb K}_g^{{\rm R},\, C_A\,T_R}[G_q,G_g]$ is not
  necessary but it could be performed (i.e. adding the same integrand
  with $z\leftrightarrow 1-z$ and dividing by 2) to mirror the
  iteration of the NLL kernel.}
\begin{align}\label{eq:KDCCATR}
  {\mathbb K}_g^{{\rm DC}, \, C_A\,T_R}&[G_q,G_g] =
\int_t^{t_0} d t_{1,23} \,d t_{2,3}\int_{z_0}^{1-z_0} d z \,d z_p
P_{gg}(z) P_{q g}(z_p) \,\bigg[\bigg\{G_{q}(x\,z_p\,(1-z),t_{2,3})\notag\\
&\times G_{q}(x\, (1-z_p)\,(1-z), t_{2,3}) G_g(x\, z, t_{1,23}) -
G_{g}(x\, (1-z), t_{1,23}) \,G_g(x\, z,t_{1,23})
\bigg\}\notag\\&\times\frac{\Delta_g(t)}{\Delta_g(t_{2,3})}\,\Theta(t_{2,3}
-t_{1,23}) +\{z\leftrightarrow 1-z\}\bigg]\,,
\end{align}
 \begin{align} {\mathbb K}_g^{{\rm DC}, \, C_F\,T_R}&[G_q,G_g]
=\int_t^{t_0} d t_{2,13} \,d t_{1,3}\int_{z_0}^{1-z_0} d z \,d z_p
P_{qg}(z) P_{qq}(z_p) \,\bigg[ \bigg\{G_q(x\,(1-z),t_{2,13}) \notag\\ &\times G_g(x\,
z \,(1-z_p), t_{1,3}) G_q(x\, z\,z_p, t_{1,3}) -
G_{q}(x\, (1-z), t_{2,13}) \,G_q(x\, z,t_{2,13})
\bigg\}\notag\\ &\times\frac{\Delta_g(t)}{\Delta_g(t_{1,3})}\,\Theta(t_{1,3}
-t_{2,13})+\{2\leftrightarrow 3,\,z\leftrightarrow 1-z\}\bigg]\,,
\end{align}
\begin{align}\label{eq:KDCCA2}
  {\mathbb K}_g^{{\rm DC}, \, C_A^2}&[G_q,G_g] =
\int_t^{t_0} d t_{12,3} \,d t_{1,2}\int_{z_0}^{1-z_0} d z \,d z_p
P_{gg}(z) P_{g g}(z_p) \,\bigg[\bigg\{G_{g}(x\,z_p\,(1-z),t_{1,2})\notag\\
&\times G_{g}(x\, (1-z_p)\,(1-z), t_{1,2}) G_g(x\, z, t_{12,3}) -
G_{g}(x\, (1-z), t_{12,3}) \,G_g(x\, z,t_{12,3})
\bigg\}\notag\\&\times\frac{\Delta_g(t)}{\Delta_g(t_{1,2})}\,\Theta(t_{1,2}
-t_{12,3}) +\{z\leftrightarrow 1-z\}\bigg]\,.
\end{align}
We note that in Eqs.~\eqref{eq:KDCCATR} and~\eqref{eq:KDCCA2} the
symmetrisation in $z\leftrightarrow 1-z$ is not accompanied by a
factor of $1/2!$ since this is already included in the definition of
the leading-order splitting function $P_{gg}(z)$.

\section{Derivation of $\frac{\theta^2}{\sigma_0}\frac{d^2\sigma}{d\theta^2d z} $}
\label{app:SDquark}
In this appendix we derive the double-differential distribution in
Eq.~\eqref{eq:theta-z-dist}. It is instructive to start with the NLL
case, for which the distribution
$\frac{\theta^2}{\sigma_0}\frac{d^2\sigma}{d\theta^2d z} $ is defined
as (cf. Eq.~\eqref{eq:dsigma})
\begin{equation}\label{eq:dsigma-quark}
\frac{\theta^2}{\sigma_0}\frac{d^2\sigma^{(f)}}{d\theta^2d z} =
\sum_{m=1}^{\infty} \int \,dP^{(f)}_m\,\theta^2\delta(\theta^2-\theta^2_{\rm pass})\delta(z - z_{\rm pass})\,,
\end{equation}
where $\theta_{\rm pass}$ and $z_{\rm pass}$ denote the angle and momentum
fraction of the largest-angle branching with $1-z_{\rm cut} > z_{\rm pass} > z_{\rm
  cut}$.
The probabilities $dP^{(f)}_m$ are computed with
Eq.~\eqref{eq:prob-quark} using the NLL evolution
equation~\eqref{eq:GF-NLL}.
For the sake of establishing our argument, we consider the simpler
case of the fragmentation of a quark.
For a given number $m$ of final state particles, the observable is
defined by considering all sequential primary declusterings with
respect to the hard leg that initiates the fragmentation in an angular
ordered picture, and setting $\theta$ and $z$ to the first one that
passes the $z_{\rm cut}$ (with $z_{\rm cut}\ll 1$) condition. At NLL,
each declustering amounts to a single primary emission off the initial
hard leg, which then fragments inclusively. Therefore, we can ignore
the secondary branchings of the gluons and approximate their
generating functional with the first order expansion, that is
\begin{equation}
G_g(x,t) \simeq u\,.
\end{equation}
With this simplification, the first few terms of the series read
\begin{align}
  \int dP^{(q)}_1 &= \Delta_q(t)\,,\\
  \int dP^{(q)}_2 &= \Delta_q(t)\,\int_{t}^{t_0}d t_1\int_{z_0}^{1-z_0} d
  z_1\,P_{qq}(z_1)\,,\\
  \int dP^{(q)}_3 &= \Delta_q(t)\,\int_{t}^{t_0}d t_1 \int_{t_1}^{t_0}d t_2\int_{z_0}^{1-z_0} d
               z_1 d z_2\,P_{qq}(z_1)\,P_{qq}(z_2)\,,\\
  ... &= ...
\end{align}
The final state with a single parton does not contribute to the
distribution (more precisely it contributes a $\delta(\theta^2)$
term), while the final state with two partons (described by
$dP_2^{(q)}$) contributes according to the measurement function
\begin{equation}
\theta^2\delta(\theta^2-\theta^2_1)\delta(z-z_1)\Theta(z_1 - z_{\rm cut})\,.
\end{equation}
Similarly, the state with three partons will involve the measurement
function
\begin{equation}
\theta^2\left(\delta(\theta^2-\theta^2_1)\delta(z-z_1)\Theta(z_1 - z_{\rm cut})+\delta(\theta^2-\theta^2_2)\delta(z-z_2)\Theta(z_{\rm cut}-z_1) \Theta(z_2-z_{\rm cut})\right)\,,
\end{equation}
where the first term corresponds to a configuration in which the
first, largest-angle emission (declustering) passes the $z_{\rm cut}$
condition, while the second term corresponds to a configuration in
which the first emission fails the condition and the observable is
then defined by the second branching provided it passes the cut. The
configuration in which neither of the declusterings passes the cut
only gives a $\delta(\theta^2)$ contribution to the observable, and
can be discarded for the differential distribution.

One can then evaluate explicitly the infinite sum in
Eq.~\eqref{eq:dsigma-quark} by factoring out the contribution of the
branching that passes the cut and summing inclusively over the
remaining branchings. If we trade the angular ordering for a $1/n!$
combinatorial factor, we obtain the formula
\begin{align}
  \frac{\theta^2}{\sigma_0}\frac{d^2\sigma^{(f)}}{d\theta^2d z} &=
  \Delta_q(t)\, \frac{\alpha_s(E^2g(z)^2\theta^2)}{2\pi}\,P_{qq}(z)
  \notag\\
  &\times \exp\left\{\int_{t}^{t_0}d t'\int_{z_0}^{1-z_0} d
  z\,P_{qq}(z)\, \left(\Theta(z_{\rm cut}-z)+\Theta(\theta-\theta')\Theta(z - z_{\rm cut})\right)\right\}\, \,.
\end{align}
We can now evaluate the integral in the exponent and combine it with
the Sudakov $\Delta_q(t)$ given in Eq.~\eqref{eq:sudakov-nll}. After
taking the limits $t_0\to \infty$ and $z_0\to 0$ we obtain
\begin{equation}\label{eq:dsigma-nll}
  \frac{\theta^2}{\sigma_0}\frac{d^2\sigma^{(f)}}{d\theta^2d z} =
  \frac{\alpha_s(E^2g(z)^2\theta^2)}{2\pi}\,P_{qq}(z)
  \,e^{-R^{\rm NLL}_q(\theta,z_{\rm cut}) }\,,
\end{equation}
where $t_\theta$ denotes the evolution time corresponding to the angle
$\theta$ and
\begin{equation}
  R^{\rm NLL}_q(\theta,z_{\rm cut}) \equiv \int_{0}^{t_\theta}d t'\int_{z_{\rm
    cut}}^{1-z_{\rm
    cut}} d
  z\,P_{qq}(z)
\end{equation}

The quantity in the exponent defines the Sudakov radiator featuring in
Eq.~\eqref{eq:theta-z-dist}.
The NNLL version of Eq.~\eqref{eq:dsigma-nll} can be obtained
following the same procedure. We start from Eq.~\eqref{eq:dsigma},
where now the matching coefficient needs to be evaluated at one loop,
that is
\begin{equation}\label{eq:c1q}
  C(\alpha_s) = 1+\frac{\alpha_s(E^2)}{2\pi} C_q^{(1)}+{\cal
  O}(\alpha_s^2)\,.
\end{equation}
The constant $C_q^{(1)}$ contains hard-virtual corrections to the Born
process, as well as non-logarithmic terms of soft and/or collinear
origin that contribute in the limit
$\theta\to 0$.\footnote{\label{foot:scheme}Importantly, the coupling in
  Eq.~\eqref{eq:c1q} is evaluated at the hard scale of the process
  (i.e. $\mu^2\sim E^2$), which formally corresponds to working in a
  resummation scheme~\cite{Catani:2000vq} in which such
  non-logarithmic terms are evaluated at the hard scale. In an
  alternative resummation scheme one may want to calculate some of
  these non-logarithmic terms (for instance those of soft and/or
  collinear origin) at the low scales (soft or collinear) of the
  problem, so that their running generates logarithmic terms at higher
  perturbative orders. Consistently with a resummation scheme
  transformation, this implies that the higher-order logarithmic terms
  generated by the running of the coupling must be subtracted from the
  anomalous dimension $ {\mathcal B}^f_2(z)$ computed here.}

As a second step, we consider the functional derivatives of the NNLL
quark generating functional, starting from the term in the first line
of Eq.~\eqref{eq:Kvquark}. In doing so, we can work with the
approximation $E_p\equiv E$ in the evolution time~\eqref{eq:time},
since we work under the assumption that $z_{\rm cut}\ll 1$. The effect
of this term is simply to replace $P_{qq}$ in
Eq.~\eqref{eq:dsigma-nll} with the full inclusive emission probability
${\mathcal P}_{q}(z,\theta)$ given in
Eqs.~\eqref{eq:Kvquark},~\eqref{eq:quarkP}.
We then consider the contribution of the ${\mathbb K}_q^{\rm finite}$
term, given in Appendix~\ref{app:kernels}. By inspecting the structure
of ${\mathbb K}_q^{\rm finite}$ we observe that the local counterterms
in Eqs.~\eqref{eq:KR},~\eqref{eq:KDC} are defined precisely by means
of the same projection between the $1\to 3$ and the $1\to 2$ phase
spaces that was used in the definition and calculation of
$ {\mathcal B}^f_2(z)$. Therefore the observable is defined in exactly
the same way for the real contributions as well as for the
corresponding counterterms, giving a zero correction. Therefore, the
contribution of ${\mathbb K}_q^{\rm finite}$ to the differential
distribution
$\frac{\theta^2}{\sigma_0}\frac{d^2\sigma^{(f)}}{d\theta^2d z} $
vanishes trivially, and one is left with the general form in
Eq.~\eqref{eq:theta-z-dist}.
The same derivation holds for gluon jets, leading to the expression
\begin{equation}\label{eq:theta-z-dist}
    \frac{\theta^2}{\sigma_0}\frac{d^2\sigma}{d\theta^2d z} =
    \left(1+\frac{\alpha_s(E^2)}{2\pi} C_g^{(1)}\right) e^{-R^{\rm NNLL}_g(\theta,z_{\rm
          cut})}\,\frac{\alpha_s(E^2g(z)^2\theta^2)}{2\pi}\,\left({\cal P}_{gg}(z,\theta)+{\cal P}_{qg}(z,\theta)\right)\,,
\end{equation}
with
\begin{equation}
  R^{\rm NNLL}_g(\theta,z_{\rm cut}) \equiv \int_{0}^{t_\theta}d t'\int_{z_{\rm
    cut}}^{1-z_{\rm
    cut}} d
  z\, \left({\cal P}_{gg}(z,\theta')+{\cal P}_{qg}(z,\theta')\right)\,,
\end{equation}
where the inclusive emission probabilities are given in Eq.~\eqref{eq:gluonP}.
In the extraction of $ {\mathcal B}^g_2(z)$ we have chosen to work
with the fragmentation of a gluon jet defined as one hemisphere in the
two-jet limit of the $H\to gg$ decay in the heavy-top-mass limit. In
this case, the one-loop matching coefficient is given by%
\begin{equation}
C_g^{(1)} = C_A\left(4 \ln 2 \ln \zc+\frac{67}{9}-\frac{\pi^2}{3}\right)-T_R n_f \frac{23}{9}\,.
\end{equation}

\section{Double-soft end-point contributions to ${\cal B}_2^g(z)$}
\label{app:clustering}
In this appendix we discuss how to devise an alternative scheme for
$ {\mathcal B}^{g,C_A^2}_2(z)$ in which the clustering and endpoint
contributions, whose origin lies in the double-soft limit, completely
disappear. The emergence of the clustering correction in
Eq.~\eqref{eq:expform-gluon}, and likewise the end-point contribution
in Eq.~\eqref{eq:B2ofzca2}, is due to the specific definition of $z$
and $\theta$ in the kinematic projection that we used to define
$ {\mathcal B}^{g,C_A^2}_2(z)$.  To make the physics clear it is
useful to recapitulate the case of quark jets~\cite{Dasgupta:2021hbh},
in which the definition of $ {\mathcal B}^q_2(z)$ does not involve
such terms. The procedure given in Sec.~\ref{sect:defs} is based upon
decomposing, for each colour channel, the three-particle phase space
into angular sectors each containing exactly one collinear
singularity. In the case of quark jets, this task is transparent. For
correlated channels, there is a unique collinear singularity between
the offspring of the gluon decay, and thus there is a single sector
defined according to our procedure.
For the $C_F^2$
channel the only collinear singularities arise when emissions $g_1$ or
$g_2$ are collinear to the quark $q_3$, and thus there are two sectors in our procedure.
These features lead to a natural definition of $z$ and $\theta$, in all colour channels, that is based
on a \textit{naive} clustering procedure in which pairs of partons which do not develop a collinear singularity never get clustered together~\cite{Dasgupta:2021hbh}.
An equally valid, albeit tedious, procedure for defining $z$ and
$\theta$ would be to follow a strict C/A algorithm and define $z$ as
the momentum fraction between the two branches produced by the C/A
declustering. This procedure would differ from the previous one in
finite angular regions resulting in an alternative scheme for
$ {\mathcal B}^q_2(z)$, which would contain a clustering correction.

Let us now move to the case of gluon jets.  As discussed in the main
text, the $C_A^2$ channel has a richer structure of collinear
singularities which involves all three partons (gluons) on an equal
footing.
In particular, it is not possible to separate out correlated and independent emissions. 
In this case one cannot use the naive clustering argument adopted for
quark jets and for this reason we had to resort to strict C/A procedure
outlined in the article. This ultimately leads to the appearance of
the clustering correction in Eq.~\eqref{eq:expform-gluon}.
However, in a situation in which two of the three final state gluons
are soft the parent gluon can be uniquely identified and one can
separate the double-soft squared amplitude into the sum of
correlated and independent emission terms which are identical to those
in the quark case up to Casimir scaling. 

One could then introduce an alternative scheme in which the
double-soft limit is treated as in the quark case, while the
hard-collinear leftover is treated as explained in
Sec.~\ref{sect:defs}, i.e. using strict C/A declustering.
Given that the clustering correction originates from the double-soft
region of phase space, this alternative $ {\mathcal B}^g_2(z)$ will
directly integrate to
\begin{equation}\label{eq:mod-scheme}
\int_0^1d z \, {\mathcal B}^g_2(z) = -\gamma^{(2)}_g+b_0 X_\theta^2\,.
\end{equation}
A scheme of this type has the advantage of eliminating the end-point contribution present in Eq.~\eqref{eq:B2ofzca2}, but 
presents a slight disadvantage in
that it makes the calculation of ${\mathbb K}_q^{\rm finite}$ more
cumbersome due to the different projections adopted in the double-soft
and in the hard-collinear limits.

Below we present the quantity $ {\mathcal B}^g_2(z)$ in this alternate
scheme, which differs from the one given in the article exclusively in
the $C_A^2$ colour channel.
In order to carry out the computation of $\mathcal{B}_2^{g,C_A^2}(z)$
in the new scheme, we start by separating the double-soft limit of the
splitting function, which satisfies Casimir scaling and is
therefore identical (up to an overall quadratic Casimir operator) to
the quark case.
Let gluons $(g_i,g_j)$ be soft, i.e. $z_i,z_j \ll 1$, while parton $k$
is hard $z_k \to 1$. In the triple-collinear limit we have
\begin{align}\label{eq:DSdef}
	\langle \hat{P}_{g_i  g_j; p_k } \rangle = C_k^2  \langle \hat{P}^{\rm ind.}_{g_i  g_j; p_k } \rangle + C_k C_A \langle \hat{P}^{\rm corr.}_{g_i  g_j; p_k } \rangle \ ,
\end{align}
where $C_k$ is the Casimir of the hard parton $k$. The various functions read~\footnote{Notice here in particular that we do not send $z_k \to 1$ in the
	double-soft splitting functions, because the triple-collinear phase
	space will be exactly retained when the computation is performed. This choice is made to ensure the subtraction in Eq.~\eqref{eq:subsplit} is achieved locally in phase space.}
\begin{align}
	\langle \hat{P}^{\rm ind.}_{g_i  g_j; p_k } \rangle = \frac{4}{s_{ik} s_{jk}} \frac{z_k}{z_i z_j} \ ,
\end{align}
and
\begin{align}\label{eq:corrfunction}
	\nn
	\langle \hat{P}^{\rm corr.}_{g_i  g_j; p_k } \rangle  &= \frac{(1-\ep)}{4 z_k (s_{ik} + s_{jk})^2 s_{ij}^2 } \left(2 \frac{z_i s_{jk}  - z_j s_{ik}}{z_i + z_j}\right)^2 +  \frac{1}{s_{ij} s_{ik}} \left(\frac{1}{z_j} + \frac{1}{z_i + z_j}\right)- \frac{1}{2 s_{ik} s_{jk} } \frac{z_k}{z_i z_j} \\
	&+\frac{1}{2(s_{ik} + s_{jk} ) s_{ij}} \left(\frac{1}{z_i}+ \frac{1}{z_j} - \frac{8}{z_i +z_j}\right) 
	- \frac{1}{(s_{ik} + s_{jk}) s_{ik}} \frac{z_k}{z_i (z_i+z_j)} +( i \leftrightarrow j) \ .
\end{align}
%

%
In the case of three identical gluons, we have to properly symmetrise
the double-soft function in order to account for any pair of gluons
becoming soft. Therefore we define the following \textit{subtracted}
splitting function
\begin{align}\label{eq:subsplit}
	\langle  \hat{P}^{\rm sub.}_{g_1  g_2 g_3 } \rangle \equiv \frac{1}{s_{123}^2} \langle \hat{P}_{g_1  g_2 g_3 } \rangle - \langle \hat{P}_{g_1  g_2; g_3 } \rangle - \langle \hat{P}_{g_1  g_3; g_2 } \rangle - \langle \hat{P}_{g_2  g_3; g_1 } \rangle \ \ .
\end{align}
Having removed the double-soft limit, we are now ready to repeat our
standard computations using Eq.~\eqref{eq:subsplit} as our integrand,
for which we follow the sectorisation introduced in
Sec.~\ref{sect:defs}.

\begin{figure}[htbp]
	\centering
	\includegraphics[width=0.7\textwidth]{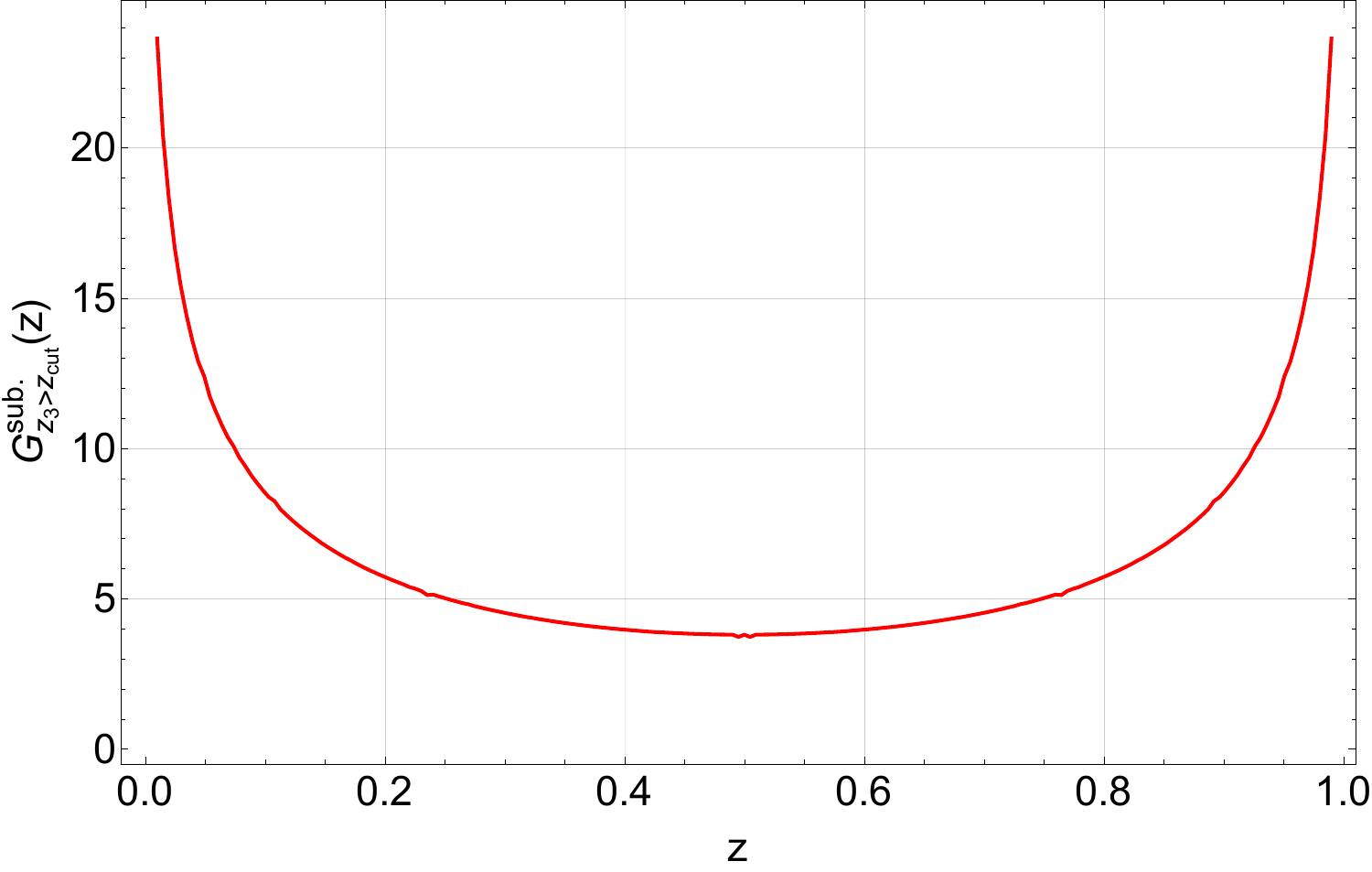}
	\caption{Plot of the function
          $G^{\text{sub.}}_{z_3 > z_\text{cut}} (z)$.}
	\label{fig:Rsymm}
\end{figure}

The results presented in Sec.~\ref{sect:ca2} are then modified as
follows:
\begin{multline}\label{eq:subpass}
  \left(\frac{\theta^2}{\sigma_0}
    \frac{\sd^2\sigma_{\mathcal{R}}^{(2)}}{\sd\theta^2 \, \sd
      z}\right)_{z_3 > z_\text{cut}}^{C_A^2,\,{\rm sub.}} =
\left(\frac{C_A \alpha_s}{2\pi}\right)^2  \left( z (1-z) \theta^2\right)^{-2\epsilon} \bigg( H^{\text{sub.}}_{\text{soft-coll.}}(z,\epsilon)
	+  H^{\text{sub.}}_{\text{coll.}}(z,\epsilon) \\
	+  H^{\text{sub.}}_{\text{soft}}(z,\epsilon)+  H^{C_A^2,\,\text{sub.}}_{\text{fin.}}(z)\bigg) \ ,
\end{multline}
where
\begin{align}
\begin{split}
		H^{\text{sub.}}_{\text{soft-coll.}}(z,\epsilon) =& 4^{\epsilon} \, \left(z^{-2\epsilon} + (1-z)^{-2\epsilon}\right) \, \left(-2 + z(1-z) \right)  \left(\frac{1}{\epsilon^2} +\frac{2 \ln 2}{\epsilon} + 4 \ln^2 2 - \frac{\pi^2}{2}\right) \ ,
			\\
		H^{\text{sub.}}_{\text{coll.}}(z,\epsilon)  = \, & 4^\ep\, p_{gg}(z)\, \left(z^{-2\epsilon} + (1-z)^{-2\epsilon}\right)  \bigg[ \frac{ (11-12 \ln 2)}{6 \epsilon } + \, \left(\frac{67}{9}-\frac{\pi^2}{3} - 4 \ln ^2 2\right) \bigg] \\ 
		&- \frac{11}{6 \ep} 4^{\epsilon} \bigg(\frac{(1-z)^{-2\ep} }{1-z} + \frac{z^{-2\ep}}{z}\bigg) +\,  4^{\epsilon } \left(z^{-2\epsilon} + (1-z)^{-2\epsilon}\right) \frac{ 2 \ln 2}{z (1-z)
			\epsilon }\\
		&- \frac{1}{z(1-z)} \bigg(\frac{67}{9} - \frac{2\pi^2}{3} - 8 \ln^2 2 \bigg)\, , \\
		H^{\text{sub.}}_{\text{soft}}(z,\epsilon) =& - \left(z^{-2\epsilon} + (1-z)^{-2\epsilon}\right) (-2+z(1-z)) \bigg(\frac{2 \ln (2)}{\epsilon } - \frac{h^{\text{sub.}}_{\rm pass}}{2}\bigg) \, ,
\end{split}
\end{align}
where the constant $h^{\text{sub.}}_{\rm soft}$ is the result of a
2-fold integration and reads
\begin{align}
h^{\text{sub.}}_{\rm soft} \simeq -6.49651797(5) \, .
\end{align}
The function $H^{C_A^2,\,\text{sub.}}_{\text{fin.}} $ has the usual
decomposition~\eqref{eq:Gz}
\begin{align}
	H^{C_A^2,\,\text{sub.}}_{\text{fin.}} (z) = -\frac{11 \ln (2)}{3}\frac{1}{ z (1-z)} + G^{\text{sub.}}_{z_3 > z_\text{cut}}(z )\, , 
\end{align}
and the remainder function $ G^{\text{sub.}}_{z_3 > z_\text{cut}}$, see Fig.~\ref{fig:Rsymm}, is
integrable in $z \in [0,1]$ and its integral reads
\begin{align}
	\int_0^1 dz \, G^{\text{sub.}}_{z_3 > z_\text{cut}} (z) \simeq 6.974 \pm 0.001 \,.
\end{align}
Similarly, below the soft threshold $z_3 < z_{\rm cut}$ we have
\begin{multline}\label{eq:subfail}
		\left(\frac{\theta^2}{\sigma_0}
          \frac{\sd^2\sigma_{\mathcal{R}}^{(2)}}{\sd\theta^2 \, \sd
            z}\right) _{z_3 < z_\text{cut}}^{C_A^2,\,\text{sub.}} =
	 \left(\frac{C_A \alpha_s}{2\pi}\right)^2 \left(-2+z(1-z)\right) \Theta\left( z- z_{\rm cut} \right) \Theta\left( 1-z_{\rm cut}  - z\right) \times \\
	 \times 4^{-\epsilon} (z(1-z))^{-2\epsilon} \, (\theta^2)^{-\epsilon} \,\ \bigg(\frac{z^{-2\epsilon}_{\rm cut}}{-\epsilon} \, \ln\frac{4}{\theta^2} -  \frac{\pi^2}{6} - \frac{1}{2} \ln^2\frac{4}{\theta^2} + h^{p_{gg}}_{\rm fail}\bigg)  \, ,
\end{multline}
where the constant $h^{p_{gg}}_{\rm fail}$ is given in
Eq.~\eqref{eq:fail-const}. The most important feature of eq.~\eqref{eq:subfail} is that its $z$-dependence is fully regular over $z \in [0,1]$ unlike Eq.~\eqref{eq:failca2B} which contains an end-point contribution.

Finally we compute the pure double-soft contributions, which can be read off from the analytic results of Ref.~\cite{Dasgupta:2021hbh} by enforcing the appropriate limit. Starting with the correlated portion of the double-soft function, eq.~\eqref{eq:corrfunction}, we find
\begin{align}\label{eq:altcorr}
\nn
		\left(\frac{\theta^2}{\sigma_0}
\frac{\sd^2\sigma_{\mathcal{R}}^{(2)}}{\sd\theta^2 \, \sd
	z}\right)^{C_A^2,\,\text{corr.}} =& \left(\frac{C_A \alpha_s}{2\pi}\right)^2 \left(z(1-z) \theta^2\right)^{-2\epsilon} \left(\frac{1}{\epsilon ^2}+\frac{11}{6 \epsilon }-\frac{2\pi
^2}{3}+\frac{67}{18}\right)  \times\\
&\times \left(\frac{(1-z)^{-2\epsilon}}{1-z} +\frac{z^{-2\epsilon}}{z}\right) \Theta\left( z- z_{\rm cut} \right) \Theta\left( 1-z_{\rm cut} - z \right) 	\, , 
\end{align}
while for the independent emission contribution we get\footnote{The third line of Eq.~\eqref{eq:altind} emerges when the gluon at larger angle fails the $z_{\rm cut}$ condition. Although this contribution can not be extracted directly from Ref.~\cite{Dasgupta:2021hbh}, it entails a straightforward computation.}
\begin{align}\label{eq:altind}
	\nn
		\left(\frac{\theta^2}{\sigma_0}
\frac{\sd^2\sigma_{\mathcal{R}}^{(2)}}{\sd\theta^2 \, \sd
	z}\right)^{C_A^2,\,\text{ind.}}  =& \left(\frac{C_A \alpha_s}{2\pi}\right)^2 \left(z (1-z)  \theta^2\right)^{-2\epsilon} \left(\frac{1}{\epsilon^2} - \frac{5\pi^2}{6}\right) \left(\frac{ z^{-2\epsilon}}{1-z} +\frac{(1-z)^{-2\epsilon}}{z}\right)\times \\ \nn
	&\times \Theta\left( z- z_{\rm cut}\right) \Theta\left( 1 -z_{\rm cut} - z\right) \\ \nn
   +&  \left(\frac{C_A \alpha_s}{2\pi}\right)^2  \frac{1}{z(1-z)}\Theta\left( z - z_{\rm cut} \right) \Theta\left( 1-z_{\rm cut} - z\right) \times \\
   &\times 4^{-\epsilon} (z(1-z))^{-2\epsilon} \, (\theta^2)^{-\epsilon} \,\ \bigg(\frac{z^{-2\epsilon}_{\rm cut}}{-\epsilon} \, \ln\frac{4}{\theta^2} -  \frac{\pi^2}{6} - \frac{1}{2} \ln^2\frac{4}{\theta^2} \bigg) .
\end{align}
Adding all the results in Eqs.~\eqref{eq:subpass}, \eqref{eq:subfail},
\eqref{eq:altcorr}, \eqref{eq:altind}, and the relevant virtual
corrections, the alternative scheme yields
\begin{align}
	\mathcal{B}_2^{g,C_A^2,\,\text{alt.}}\left(z \right) =  G^{\text{sub.}}_{z_3 > z_\text{cut}} (z) + \mathcal{B}^{g,\,C_A^2,\,\rm{analytic,\,alt.}}_{2}\left(z\right) \, ,
\end{align}
where
\begin{align}\label{eq:analb2base}
	\nn
	\mathcal{B}^{g,\,C_A^2,\,\rm{analytic,\,alt.}}_{2}\left(z \right) &= 
	\frac{1}{18}\left(2-z(1-z)\right) \left(-18 \, h^{\text{sub.}}_{\rm soft}- 18 \, h^{p_{gg}}_{\rm fail}+6\pi^2 -132 \ln 2 - 72 \ln^2 2 \right) \\ \nn
	&+\frac{-265+134 z - 134 z^2}{18}
	+ \frac{11 \left(1-3z +2 z^2-z^3\right) }{6(1-z)} \ln z \\ \nn
	&+ \frac{11 \left(-1 +2 z -z^2 +z^3 \right) }{6z}  \ln(1-z) - \frac{11}{6}\frac{\ln z}{1-z} - \frac{11}{6} \frac{\ln(1-z)}{z} \\
	&- 2\,  p_{gg}(z) \ln (1-z) \ln z + \frac{11}{6}(2-z(1-z)) \ln(z(1-z)) \, .
\end{align}
The integral of $\mathcal{B}_2^{g,C_A^2,\,\text{alt.}}(z)$ then reads
\begin{align}\label{eq:b2alt}
\int_0^1 dz \, \mathcal{B}_2^{g,C_A^2,\,\text{alt.}}(z) \simeq -7.858 \pm 0.001 \, .
\end{align}
It is easy to verify that eq.~\eqref{eq:b2alt} agrees with
the sum rule given in Eq.~\eqref{eq:mod-scheme}.

\section{Comparison to \texttt{Event2} for quark-jet observables}
\label{app:event2comparison}

\begin{figure}[hb!]
  \begin{subfigure}[t]{0.4825\linewidth}
  \centering
    \includegraphics[width=1\textwidth,page=1]{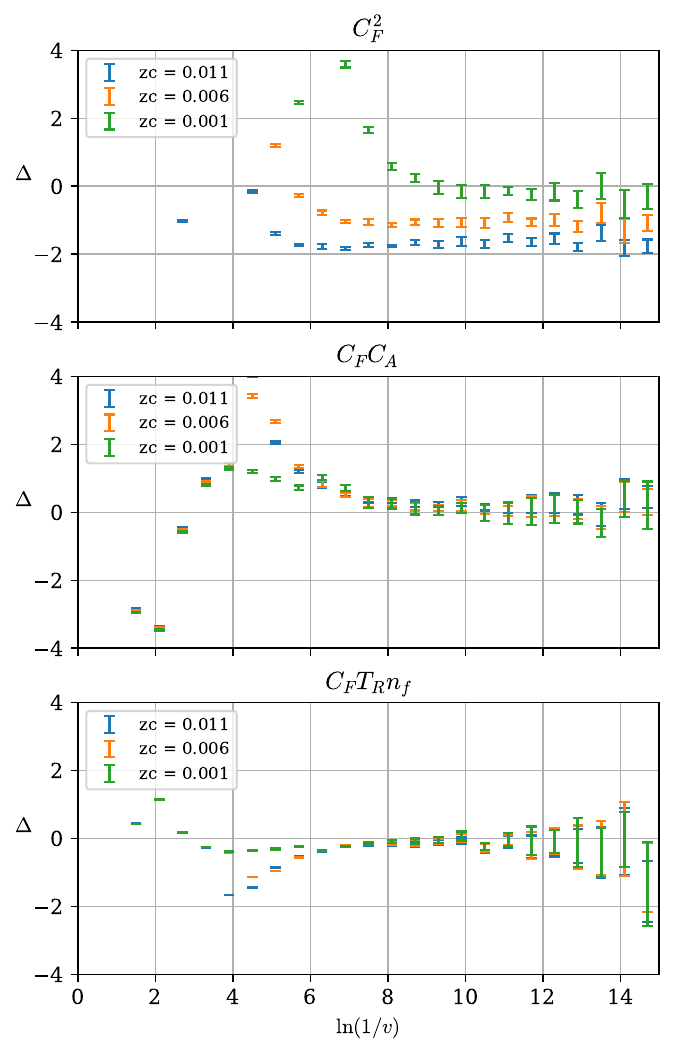}
    \caption{$\Delta_2(FC_0,\lambda_0)$}
  \end{subfigure}
   \begin{subfigure}[t]{0.49\linewidth}
   \centering
    \includegraphics[width=1\textwidth,page=2]{figures/FCvsAng_B2Check}
    \caption{$\Delta_2(FC_{1/2},\lambda_{1/2})$}
  \end{subfigure}
  \caption{Validation of resummation against \texttt{Event2} for quark jets, using three different values of $\zc$. See text for details. }\label{fig:event2_comparison}
\end{figure}

In this appendix, we present the comparison between the $\mathcal{O}(\alpha_s^2)$ expansions of our resummed results with the numerical fixed-order code \texttt{Event2}. In order to focus on the NNLL terms, and improve the precision on the pure $\mathcal{O}(\alpha_s^2)$ contribution, we define the difference between observable distributions (as done e.g.\ in~Ref.~\cite{Banfi:2014sua} for ungroomed event shapes)
\begin{align}
	\Delta_2(FC_x,\lambda_x) \equiv \frac{1}{\sigma_0} \left(\frac{d\sigma^{(2)}}{d \ln (1/FC_x)}- \frac{d\sigma^{(2)}}{d \ln (1/\lambda_x)}\right) \, ,
\end{align}
and we plot the quantity
\begin{align}
	\Delta \equiv \Delta_2\big|_{\rm Ev2}(FC_x,\lambda_x) - \Delta_2\big|_{\rm NNLL}(FC_x,\lambda_x) \, .
\end{align}
If our resummation correctly captures all NNLL terms, then we would expect the quantity $\Delta$  to tend to zero as $\ln 1/v$ grows large, in all colour channels. For groomed observables, however, the resummation we have carried out is valid in the regime $v \ll \zc \ll 1$ and hence neglects log-enhanced terms which are power-suppressed in $\zc$. Therefore, we expect to find agreement with \texttt{Event2} in the limit of $\zc \to 0$. Since taking this limit is numerically challenging, we show plots for three different values of $\zc$, showing convergence between our results and \texttt{Event2} as this limit is approached. The plots shown in Fig.~\ref{fig:event2_comparison} are obtained with $3\times10^{12}$ events, although residual statistical instabilities remain in the \texttt{Event2} results. We note that in all colour channels, the agreement with \texttt{Event2} substantially improves with decreasing $\zc$, strongly validating our resummed predictions.

\bibliographystyle{JHEP}
\bibliography{gluon}
\end{document}